\documentclass[a4paper,11pt]{article}
\usepackage{jheppub}
\usepackage{amsmath}
\usepackage{pagecolor}
\usepackage{marvosym}
\usepackage{hyperref}
\usepackage[normalem]{ulem}
\usepackage{amsthm}
\usepackage{slashed}
\usepackage{amsfonts}
\usepackage{amssymb}
\usepackage{textcomp}
\usepackage{graphicx}
\usepackage{bm}
\usepackage{verbatim}
\usepackage{graphicx,color}
\usepackage{epsfig}

\usepackage{soul}

\definecolor{orange}{rgb}{1,0.5,0}
\definecolor{col1}{RGB}{153, 52, 121}

\usepackage{amsfonts,amsmath,amssymb}

\newcommand{\cool}{\ensuremath{%
  \mathchoice{\includegraphics[height=2ex]{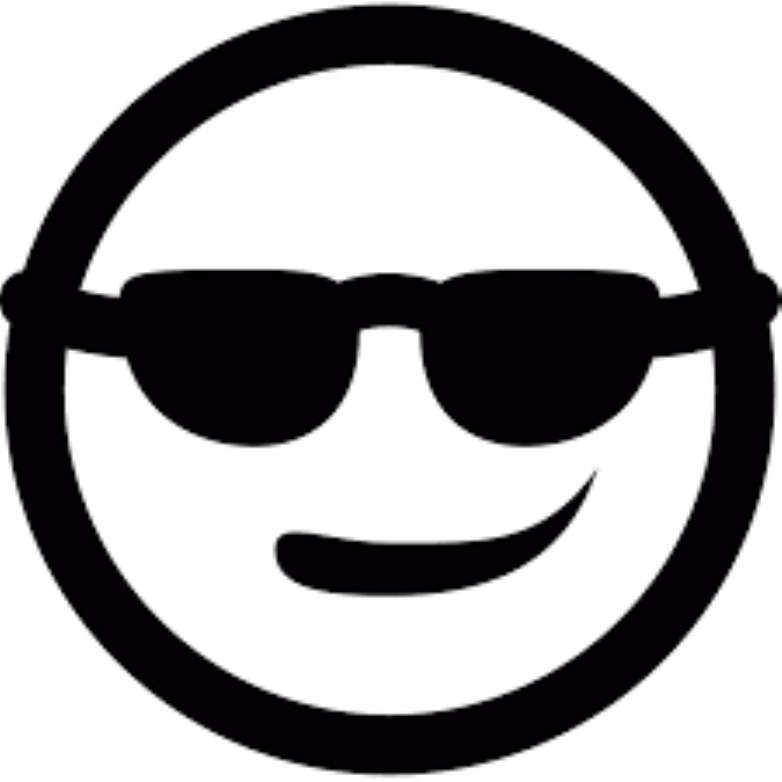}}
    {\includegraphics[height=2ex]{cool}}
    {\includegraphics[height=1.5ex]{cool}}
    {\includegraphics[height=1ex]{cool}}
}}

\newcommand{\panda}{\ensuremath{%
  \mathchoice{\includegraphics[height=3ex]{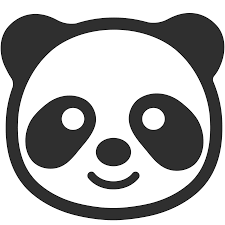}}
    {\includegraphics[height=3ex]{panda}}
    {\includegraphics[height=2.5ex]{panda}}
    {\includegraphics[height=2ex]{panda}}
}}

\theoremstyle{definition}

\begin{document}

\subheader{CCTP-2018-2\\
ITCP-IPP 2017/21}

 \title{
 \Huge \center A smeared quantum phase transition in disordered holography\color{black}}
 
 \author[a]{Martin Ammon}
 
 \author[\cool]{, Matteo Baggioli} 
  
 \author[a]{, Amadeo Jim\'enez-Alba}

  \author[a]{, Sebastian Moeckel}

 \affiliation[a]{Theoretisch-Physikalisches Institut, Friedrich-Schiller-Universit\"at Jena,
Max-Wien-Platz 1, D-07743 Jena, Germany.} 

 \affiliation[\cool]{Crete Center for Theoretical Physics, Institute for Theoretical and Computational Physics, Department of Physics, University of Crete, 71003
Heraklion, Greece.}

\emailAdd{martin.ammon@uni-jena.de}
\emailAdd{mbaggioli@physics.uoc.gr}
\emailAdd{amadeo.jimenez.alba@uni-jena.de}
\emailAdd{sebastian.moeckel@uni-jena-de}

\abstract{We study the effects of quenched one-dimensional disorder on the holographic Weyl semimetal quantum phase transition (QPT), with a particular focus on the quantum critical region. We observe the smearing of the sharp QPT linked to the appearance of \textit{rare} regions at the horizon where locally the order parameter is non-zero. We discuss the role of the disorder correlation and we compare our results to expectations from condensed matter theory at weak coupling. We analyze also the interplay of finite temperature and disorder. Within the quantum critical region we find indications for the presence of log-oscillatory structures in the order parameter hinting at the existence of an IR fixed point with discrete scale invariance.}

\maketitle

\section{Introduction}
\begin{figure}[htp]
\centering
\includegraphics[width=\textwidth]{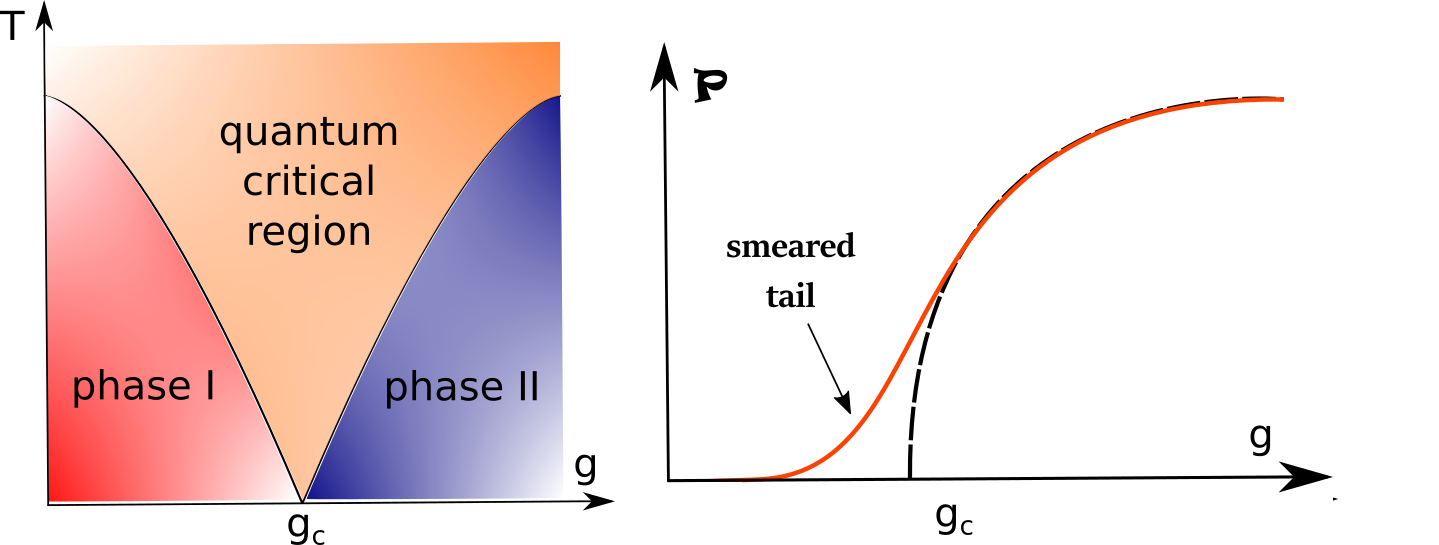}
\caption{\textbf{Left:} Sketch of the phase diagram as a function of the temperature $T$ and the external non-thermal parameter $g$. The quantum phase transition between the two phases happens at $g\equiv g_c$ and $T=0$ but it affects the physics of the system on a larger region extending towards the finite temperature regime, \textit{i.e.} the QCR (orange area). \textbf{Right:  }The evolution of the order parameter $\mathcal{P}$ across the quantum phase transition at $T=0$. The black dashed line is the clean case where the QPT is sharp while the orange line is the smeared disordered QPT.}
\label{fig1}
\end{figure}

Phase transitions are ubiquitous phenomena in nature. They play an important role in very diverse contexts; from the boiling of the water for your spaghetti, to the large scale structure of the universe, to DNA dynamics. During a phase transition the properties of a system can change abruptly, often discontinuously, upon dialing an external parameter such as temperature, pressure, or others. Differently from \textit{thermal} phase transitions, \textit{quantum} phase transitions (QPT) \cite{sachdev2011,2003RPPh...66.2069V} occur at zero temperature $T=0$ and at a precise \textit{critical} value of an external parameter $g\equiv g_c$, \textit{i.e.} the quantum critical point. Although \textit{absolute} zero temperature is clearly physically not realizable, the consequences of the QPT extend also to a finite temperature region near the critical point, known as \textit{quantum critical region} (QCR) (see fig.~\ref{fig1}). More interestingly, inside the QCR, the critical nature of the system manifests itself in unconventional but universal physical behaviors such as the linear in T scaling of the resistivity in \textit{strange metals} \cite{Bruin804}. In recent years QPTs have attracted the interest of many theorists and experimentalists especially in the condensed matter community where more and more examples are being discovered: from exotic magnetism, to high-Tc superconductivity and metal-insulator transitions. Quantum phase transitions still represent a big scientific challenge and robust and controllable theoretical models are still under investigation \cite{2007arXiv0709.0964V}. 

The presence of impurities or disorder is unavoidable in realistic situations and it can have a deep impact on the physical features of a system and its behaviour across possible phase transitions. The effects of disorder  on quantum phase transitions are much stronger than on their classical counterparts and they are subject of intense study in the present days \cite{2006JPhA...39R.143V,2013AIPC.1550..188V}.
The common lore to assess whether a phase transition is stable with respect to a small amount of \textit{randomness} relies on the so-called Harris criterion \cite{0022-3719-7-9-009}. According to that criterion, the phase transition will not be affected by the presence of disorder as far as the correlation length critical exponent $\nu$ satisfies the inequality $d\,\nu>2$, with $d$ the number of spatial dimensions in which randomness is present. 

The Harris criterion refers to the behaviour of the system at large length scales and it is not able to capture possible existing features at finite length scale. It thus represents a necessary condition for the stability of a clean critical point, but not a sufficient one. In other words disorder is defined as \textit{Harris-irrelevant} if its effects become less and less important at larger scale.  It is nowadays clear that such a criterion has to be completed and new effects, \textit{i.e.} rare regions effects, play a fundamental role in disordered QPTs \cite{2013arXiv1309.0753V,2004PSSBR.241.2118V}. In particular the generic consequence of disorder is to smear the sharp quantum phase transition (see fig.~\ref{fig1}) \cite{2008PhRvL.100x0601H,PhysRevLett.90.107202}.

The smearing or rounding of the sharp QPTs is believed to be a direct manifestation of the so-called rare regions or Griffiths effects \cite{PhysRevLett.23.17,PhysRevLett.54.1321,2010JLTP..161..299V}. First experimentally observed in quantum magnets \cite{PhysRevLett.90.107202}, these phenomena appear in a variety of systems, ranging from classical Ising magnets with planar defects \cite{0305-4470-36-43-017} to
nonequilibrium spreading transitions in the contact process \cite{PhysRevE.70.026108}. Furthermore, disorder correlations are of special relevance for QPTs; even short-range correlations can qualitatively modify quantum smeared phase transitions. Concretely positive correlations (like in the case of impurity atoms which attract each other) have been shown to enhance the smearing of the transition \cite{PhysRevLett.108.185701,2013AIPC.1550..263N,0295-5075-97-2-20007}. The understanding of the physics of disordered quantum phase transitions is just at a preliminary stage and a complete classification of the various scenarios is still lacking.\\[0.1cm]
\\
Holography, also known as \textit{gauge-gravity duality}, has emerged in the last decade as a powerful and effective framework to tackle condensed matter questions dealing with \textit{quantum criticality} and strongly correlated systems \cite{Hartnoll:2016apf,Ammon:2015wua,zaanen2015holographic}.
Quantum phase transitions have been already realized in holography in various forms \cite{DHoker:2009mmn,DHoker:2010onp,Iqbal:2010eh,Landsteiner:2015pdh,Gubankova:2014iha,Iqbal:2011aj,Donos:2012js,Baggioli:2016oqk,Baggioli:2016oju}. The idea is that, dialing an external parameter $g$ on top of an extremal $T=0$ gravitational setup, a specific order parameter $\mathcal{P}$\footnote{Note that in our case the order parameter will be the anomalous Hall conductivity and therefore given by two point function of a current operator. An analogous situation appears within the realm of metal-insulator transitions \cite{ 2011arXiv1112.6166D} which have been studied in holography for example in \cite{Donos:2012js,Baggioli:2016oqk,Baggioli:2016oju}. This is in contrast with the ``Landau-Ginzburg-Wilson'' paradigm of phase transitions where the order parameter is the vacuum expectation value of a local operator.} can emerge at a critical value $g\equiv g_c$ and induce a quantum transition between two different phases, \textit{i.e.} two qualitatively different solutions of the bulk equations of motion. As a benchmark example we will consider in this paper the holographic quantum phase transition from a topologically nontrivial Weyl semimetal to a trivial one introduced in \cite{Landsteiner:2015lsa,Landsteiner:2015pdh}.\\[0.1cm]

Weyl semimetals are materials featuring crossing of bands at isolated, non degenerate points, \textit{i.e.} Weyl nodes, in the Brillouin zone at the Fermi level \cite{Xu613,Liu864,Hosur:2013kxa,2015Sci...349..622L,2015PhRvX...5c1013L}. Crucially, the effective description of the degrees of freedom close to these points is that of relativistic fermions. This implies, among others, that anomaly effects related to chiral fermions are present in these materials.

Anomalies play a fundamental role in high energy physics \cite{Kharzeev:2015znc} and lead to concrete physical manifestations such as the decay of the neutral Pion to two photons. The observation that anomalies can also have a leading impact on condensed matter is somehow more surprising and recent. It is now well established that anomalies can be responsible for new and unexpected transport phenomena in real materials, referred as \textit{anomalous transport} (see \cite{,Landsteiner:2016led} for a review on the subject). Anomaly induced transport has been experimentally tested in Weyl and Dirac semimetals \cite{Gooth:2017mbd, Li:2014bha, PhysRevX.5.031023, nature1}.\\

\begin{figure}[htp]
\begin{center}
\includegraphics[width=0.31\textwidth]{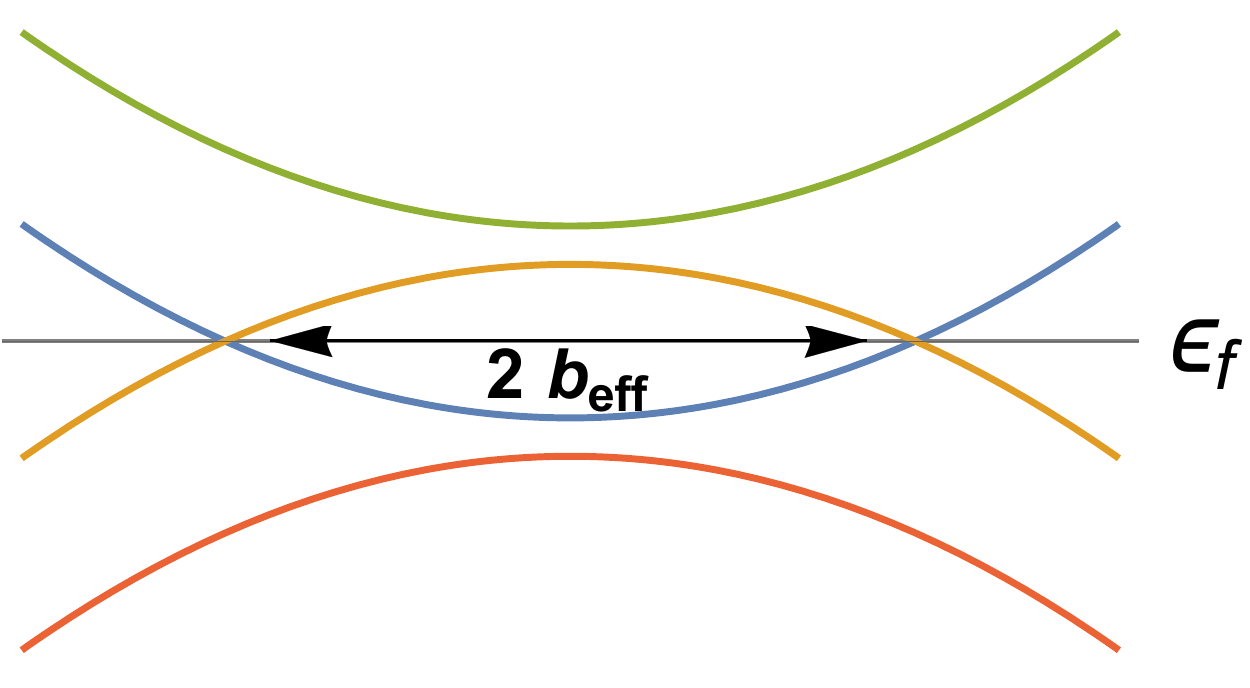}%
\quad 
\includegraphics[width=0.31\textwidth]{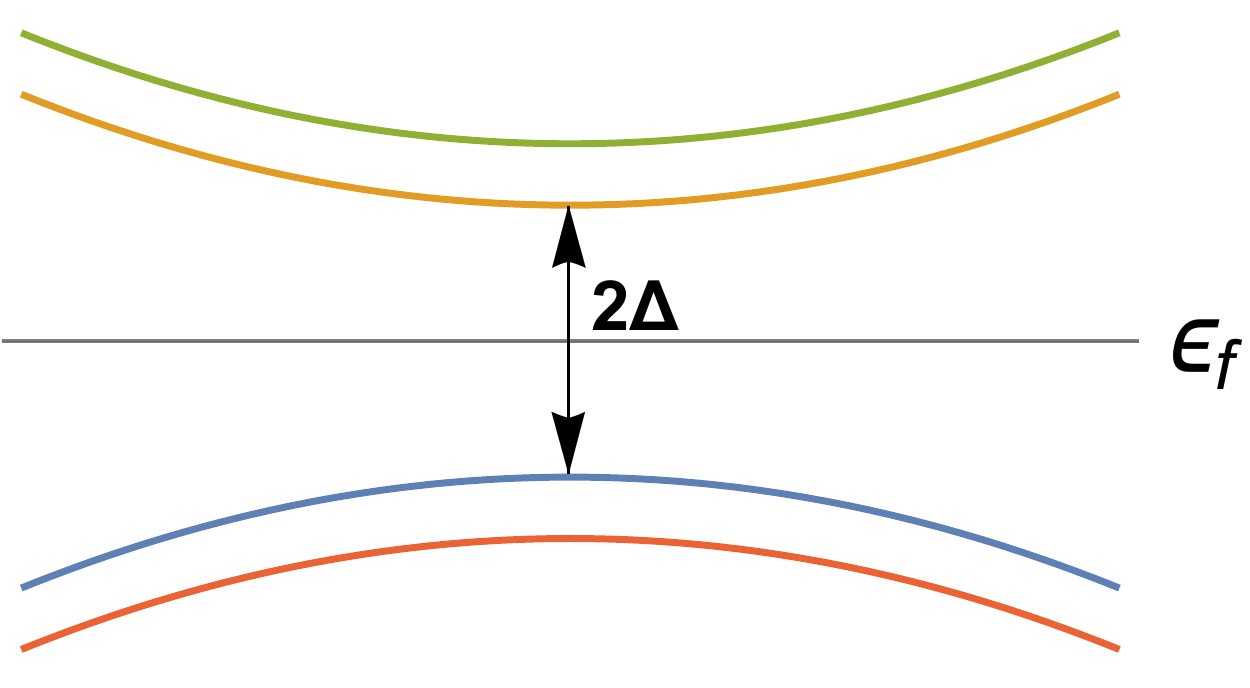}%
\quad 
\includegraphics[width=0.31\textwidth]{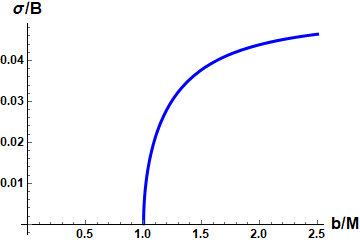}
  \caption{Sketch of the Weyl semimetal quantum phase transition. \textbf{Left: } Spectrum of the theory for $b^2>M^2$ with the two Weyl nodes. \textbf{Center :} Gapped spectrum of the theory for $b^2>M^2$. \textbf{Right :} Anomalous Hall conductivity across the QPT from the EFT result eq.\eqref{QPT}.}
  \label{fig:weyl}
\end{center}
\end{figure}

An effective low energy description of a Weyl semimetal can be realized in terms of fermions satisfying a Lorentz massive Dirac equation with a time reversal breaking parameter $b$ \cite{Grushin:2012mt}:
\begin{equation}
\left(i\,\slashed{\partial}\,-\,M\,+\,\gamma_5\,\gamma_z b\right)\,\Psi\,=\,0\,.
\end{equation}
The parameter $b$ induces a shift in the momenta of the left and right handed Weyl fermions\footnote{As a consequence of the Nielsen-Ninomiya theorem \cite{1983PhLB..130..389N} left- and
right-handed Weyl nodes in a lattice appear always in pairs.}. This model features a topological phase transition from a Weyl semimetal to a topologically trivial (massive fermions) phase. 
For $b^2>M^2$ there are two Weyl nodes in the band structure which are separated in momentum space by the effective parameter $b_{\text{eff}}=\sqrt{b^2-M^2}$. In this case the system lies in a topologically non trivial phase, \textit{i.e.} the \textit{Weyl semimetal} phase. On the contrary for 
$b^2<M^2$ the theory becomes gapped with the gap being $\Delta=\sqrt{M^2-b^2}$ and the system exhibits a topologically trivial insulating state. In other words a topological quantum phase transition appears at $b/M=1$ and exactly zero temperature. The order parameter for the QPT is the so-called anomalous Hall conductivity (AHC) defined as the Hall conductivity at zero magnetic field
\begin{equation}\label{QPT}
\sigma_{xy}(B=0)\,=\,\frac{b^2-M^2}{2\,\pi^2}\,\Theta\left(|b|\,-\,|M|\right) \, .
\end{equation}
This can be computed with field theory methods \cite{Jackiw:1999qq} and represents a clear manifestation of the axial anomaly. In fig.~\ref{fig:weyl} we provide a simple sketch of these concepts. For a more detailed explanation we refer to \cite{Landsteiner:2016led} and references therein\footnote{See also \cite{Lucas:2016omy} for an hydrodynamic description of Weyl semimetals and their transport properties.}.

The question whether and how disorder and impurities affect the physics of the Weyl semimetals has received particular attention recently. The effective field theory for disordered Weyl semimetals has been built in \cite{PhysRevLett.114.257201,PhysRevB.93.075113} and several studies regarding the phase diagram and the topological phase transition in presence of randomness have been performed \cite{2015PhRvL.115x6603C,2015PhRvL.114t6602Z,Roy:2016amv,2017PhRvB..95a4204L,PhysRevB.94.115137,2014PhRvL.113b6602S, PhysRevB.93.075113}. Despite all the effort a robust and definitive verdict is still absent.

A simple model for Weyl semimetals was recently introduced into the holographic framework in a bottom-up fashion \cite{Landsteiner:2015lsa,Landsteiner:2015pdh}. The system exhibits indeed a sharp quantum phase transition between a trivial phase to a nontrivial state dialing a non thermal external coupling encoded in the asymptotics of specific bulk fields. The order parameter for the QPT is given by the anomalous Hall conductivity. The presence of a finite anomalous hall conductivity is a direct consequence of the breaking of time reversal symmetry and an evident manifestation of anomalous transport \cite{Landsteiner:2016led}.  Several properties of the model have been investigated in various directions \cite{Landsteiner:2017lwm,Grignani:2016wyz,Landsteiner:2016stv,Copetti:2016ewq,Rogatko:2017svr,Jacobs:2015fiv,Ammon:2016mwa}. See \cite{Liu:2018bye} for a study on other topological semimetals in holography.

On the other side in the last years particular attention has been paid to the idea of incorporating disorder and its effects in the holographic picture. Throughout our work by \textit{disorder} we will mean \textit{quenched disorder}. From the (strongly coupled) quantum field theory side \cite{Aharony:2015aea} this amounts to consider ``random couplings'' whose statistical distribution does not depend on the quantum fields. In more details we have in mind a deformation of the theory:
\begin{equation}
\mathcal{S}\,=\,\mathcal{S}_0\,+\,\int d^dx\,g_i(x)\,\mathcal{O}_i\left(x\right)
\end{equation}
where the couplings $g_i(x)$ are random. From the bulk point of view this corresponds to assuming asymptotic boundary conditions for the bulk fields $\phi_i$ dual to the operators $\mathcal{O}_i$ of the type:
\begin{equation}
\lim_{\rho \rightarrow 0}\phi_i\left(\rho,x\right)\,=\,\rho^{\#}\,g_i(x)\,+\,\dots
\end{equation}
where $\#$ is a specific number depending on the conformal dimensions $\Delta_i$ of the operators $\mathcal{O}_i$ and $\dots$ are subleading corrections at the boundary $\rho=0$.  Quenched disorder has been already considered in various holographic models\footnote{The way we here consider disorder differs substantially from the ``homogeneous'' models introduced in holography for example in \cite{Vegh:2013sk,Andrade:2013gsa,Baggioli:2014roa,Alberte:2015isw,Baggioli:2015gsa}. In those cases a relaxation time for the momentum operator is effectively introduced into the dual CFT but no explicitly disordered sources appear anywhere.} with particular emphasis on its effects on transport properties \cite{Lucas:2014sba,Lucas:2014zea,Lucas:2015lna,Donos:2014yya,Garcia-Garcia:2015crx,Andrade:2017lsc}. In addition disordered geometries and fixed points have been recently studied in  \cite{Hartnoll:2014cua,Hartnoll:2014gaa}.

Furthermore, discrete scale invariance (DSI) is believed to be intimately linked to the presence of disorder \cite{SORNETTE1998239}; it manifests itself in the peculiar log-oscillatory behaviour of the thermodynamic and transport observables due to the appearance of a complex critical exponent in the vicinity of a (quantum) phase transition. The signatures of emergent (discrete) scale invariance have been already observed in disordered holographic models in \cite{Hartnoll:2015faa,Hartnoll:2015rza,Balasubramanian:2013ux,Flory:2017mal}\footnote{Discrete scale invariance has been also studied in holographic systems \textit{without} disorder in \cite{Balasubramanian:2013ux,Flory:2017mal}.}.

Finally, the effects of disorder on thermal, but not quantum, phase transitions have been already analyzed in holography in \cite{Ryu:2011vq,Fujita:2008rs,Arean:2015sqa,Arean:2014oaa,Arean:2013mta}. Concretely in \cite{Arean:2015sqa} the appearance and effects of ``rare'' regions was studied in the context of superconducting phase transitions.

Despite all the recent efforts, to the best of our knowledge, the effects of quenched disorder on \underline{quantum} phase transitions have not been investigated so far. In this direction, we aim to provide a first study of an holographic QPT and its quantum critical region in presence of disorder. In particular, we focus on one dimensional quenched disorder within the holographic Weyl semimetal QPT and tackle the following questions:\\[0.35cm]
\centerline{\textit{Will the quantum phase transition remain sharp in the presence of quenched disorder?}}\\
\centerline{\textit{If not, how will its nature get modified?}}\\[0.035cm]

\noindent The major result of this work is that indeed the holographic sharp quantum phase transition gets smeared due to the presence of quenched disorder in accordance with the condensed matter expectations \cite{2010JLTP..161..299V}.
The latter phenomenon is linked to the appearance of localized regions at the horizon where the local order parameter is non-zero.
Moreover, the effects of disorder correlation on the smearing of the QPT are consistent with the weakly coupled arguments given for example in \cite{PhysRevLett.108.185701}.
In addition, inside the quantum critical region, we find a log-oscillatory behaviour in the order parameter which we take as evidence of the emergence of a disordered fixed point displaying discrete scale invariance.

The structure of the paper is as follows: in section \ref{sec1} we introduce the model we consider; in section
\ref{sec2} we present our main results and in section \ref{sec3} we conclude with the discussion. Appendices \ref{app2} and \ref{app:numerical} are devoted to technical details about the computations and the numerical routines used to obtain the results presented in the main text.

\section{An holographic dirty quantum phase transition}\label{sec1}
In this section we present the holographic Weyl semimetal model \cite{Landsteiner:2015lsa} and the one-dimensional Gaussian quenched disorder we use to probe the quantum phase transition.
\subsection{The holographic (clean) Weyl semimetal}
The five-dimensional holographic model consists of two Abelian gauge fields, $V$ (``vector'') and $A$ (``axial'') coupled via a Chern-Simons term and a scalar field $\Phi$ charged under the axial field. The action reads as follows:
\begin{align}
\mathcal{L}=&-\frac{1}{4}H^{ab}H_{ab}-\frac{1}{4}F^{ab}F_{ab}+(D_a\Phi)^*(D^a\Phi)-\mathcal{V}(\Phi)\nonumber\\
&+\frac{\kappa}{3}\epsilon^{abcde}A_a\left(F_{bc}F_{de}+3 H_{bc}H_{de}    \right)\,,
\end{align} 
where $D_a$ is the covariant derivative specified by $D_a\equiv \partial_a-i q A_a$, $F=dA$, $H=dV$, and $\mathcal{V}(\Phi)$ is a potential which we choose to be $\mathcal{V}(\Phi)=m^2|\Phi|^2$ (see \cite{Copetti:2016ewq} for a study on more general potentials).  

We consider backgrounds of the form:
\begin{align}
 d s^2 &= \frac{1}{\rho^2} \left( - f(\rho) d t^2 + \frac{d \rho^2}{f(\rho)}  + d x^2 + d y^2 + d z^2\right), \nonumber\\
 A &= A_z(x,\rho) d z, \nonumber\\
 \Phi &= \phi(x,\rho).\label{ansatz}
\end{align}
Here, $f(\rho)=1-\rho^4$ is the emblackening factor. The horizon is located at $\rho=1$ while the boundary of the asymptotically AdS geometry is set at $\rho=0$.  For simplicity, we restrict ourselves to the probe limit which implies that the background metric is homogeneous and fixed. Note that this limits our study to the critical region induced by the zero temperature QCP. The corresponding temperature is given by:
\begin{equation}
T = \frac{- f'(1)}{4\pi} = \frac{1}{\pi} \, .
\end{equation}
The equations of motion that follow from the inhomogeneous ansatz \eqref{ansatz} read:
\begin{align}\label{eq:eqs}
0 &= -\phi  \left(q^2 \rho ^2 A_z^2+m^2\right)+\left(\rho ^2 f'(\rho )-3 \rho  f(\rho )\right) \phi^{(0,1)} +\rho ^2 f(\rho)\phi^{(0,2)}+\rho^2 \phi ^{(2,0)} \,,\\
0 &= \left(\rho ^2 f'(\rho )-\rho  f(\rho )\right) A_z^{(0,1)}+\rho ^2 f(\rho ) A_z^{(0,2)} -2 q^2 \phi^2 A_z +\rho ^2 A_z^{(2,0)}\,,\label{eq:eqs2}
\end{align}
where the partial derivatives with respect to $x$ and $\rho$ are denoted by $F^{(k,l)}=\frac{\partial^{k+l}}{\partial x^k \partial \rho^l} F$. The spatial $x$ direction is compactified to a sphere $S^1$ for numerical convenience. 

For concreteness, throughout the paper  we choose the parameters $q=1$ and $m^2=-3.$ (see \cite{Copetti:2016ewq} for a discussion on the effect of other possible choices). The expansion for the bulk fields close to the conformal boundary $\rho=0$ reads: 
\begin{equation}\label{eq:boundaryconditions}
A_z(x, \rho)\sim b(x) \, \rho^{0}\,+\,\dots\,, \qquad\phi(x,\rho)\sim M(x) \, \rho^1 +\dots\,
\end{equation}
where the dots refer to subleading terms in the limit $\rho \rightarrow 0$.  The functions $b(x)$ and $M(x)$ are to be identified as the \textit{inhomogeneous} sources for the operators dual to the bulk fields $A_z$ and $\phi$. Let us remark that even in the ``homogeneous'' case $b(x)=b$, $M(x)=M$, the introduction of a source for the $z$ component of the gauge field $A_\mu$ accounts for the explicit breaking of the $SO(3)$ symmetry of the boundary theory and it is qualitatively related to the separation of the Weyl nodes in momentum space. At finite temperature this choice of fields allows to define two dimensionless parameters in the system, which we choose to be $\bar{b}=b/M$ and $\bar{T}=T/M$.

In the $x$-independent situation, i.e. $b(x)=b$ and $M(x)=M$,  the holographic model exhibits a topological phase transition at a certain critical value of $\bar{b}$, which can be thought of as the onset of the Weyl node separation. This was first shown in the probe limit at low temperature in \cite{Landsteiner:2015lsa} and then analyzed at zero temperature in \cite{Landsteiner:2015pdh}. The order parameter is identified with the anomalous Hall conductivity (AHC), \textit{i.e.} the Hall response at zero magnetic field. For simplicity, throughout the paper we will use the simple notation $-\sigma_{xy}(B=0)\equiv \sigma$, where $\sigma_{xy}$ is obtained from the consistent-covariant currents correlator, see appendix \ref{app2}. Concretely, it was found that the onset of the order parameter happens at $\bar{b}\approx 1.4$ as reproduced here in fig.~\ref{fig:homogeneous} . The topological phase is characterized by a finite Hall conductivity, which in the homogeneous case can be read off from the behavior of the gauge field at the horizon as \cite{Landsteiner:2015lsa}:
\begin{equation}
\label{eq:AHC_hol}
\sigma=8\,\kappa \,A_z\big|_{\rho=1}\,.
\end{equation} The behaviour of the order parameter $\sigma$ close to the QPT is of the form \cite{Landsteiner:2015pdh} :
\begin{equation}
    \sigma\,\sim\,\left(\bar{b}\,-\,\bar{b}_c\right)^\Psi\,,\qquad \Psi\,\approx\,0.211
\end{equation}
which is in contrast with the ``mean field'' result $\Psi=1/2$ coming from the field theory model \cite{Landsteiner:2016led}. 

The inhomogeneous case, however, adds some difficulty to this analysis. As shown in appendix \ref{app2} only the averaged Hall conductivity $\tilde{\sigma}$ can be obtained directly from the behaviour of the gauge field at the horizon. The concrete formula reads:
\begin{equation}
\label{eq:AHC_inhom}
\tilde{\sigma}\,=\,\int_{S^1}\text{dx}\,\sigma(x)\,=\,8\,\kappa\,\int_{S^1}\text{dx}\,A_z(x)\big|_{\rho=1}\,.
\end{equation}
where $S^1$ is the circle of the compact dimension $x$ with periodicity $L$. For simplicity, we have chosen $\kappa=1/8$ throughout this paper.

\subsection{Let's get dirty}
\begin{figure}[htp]
\begin{center}
\includegraphics[width=0.45\textwidth]{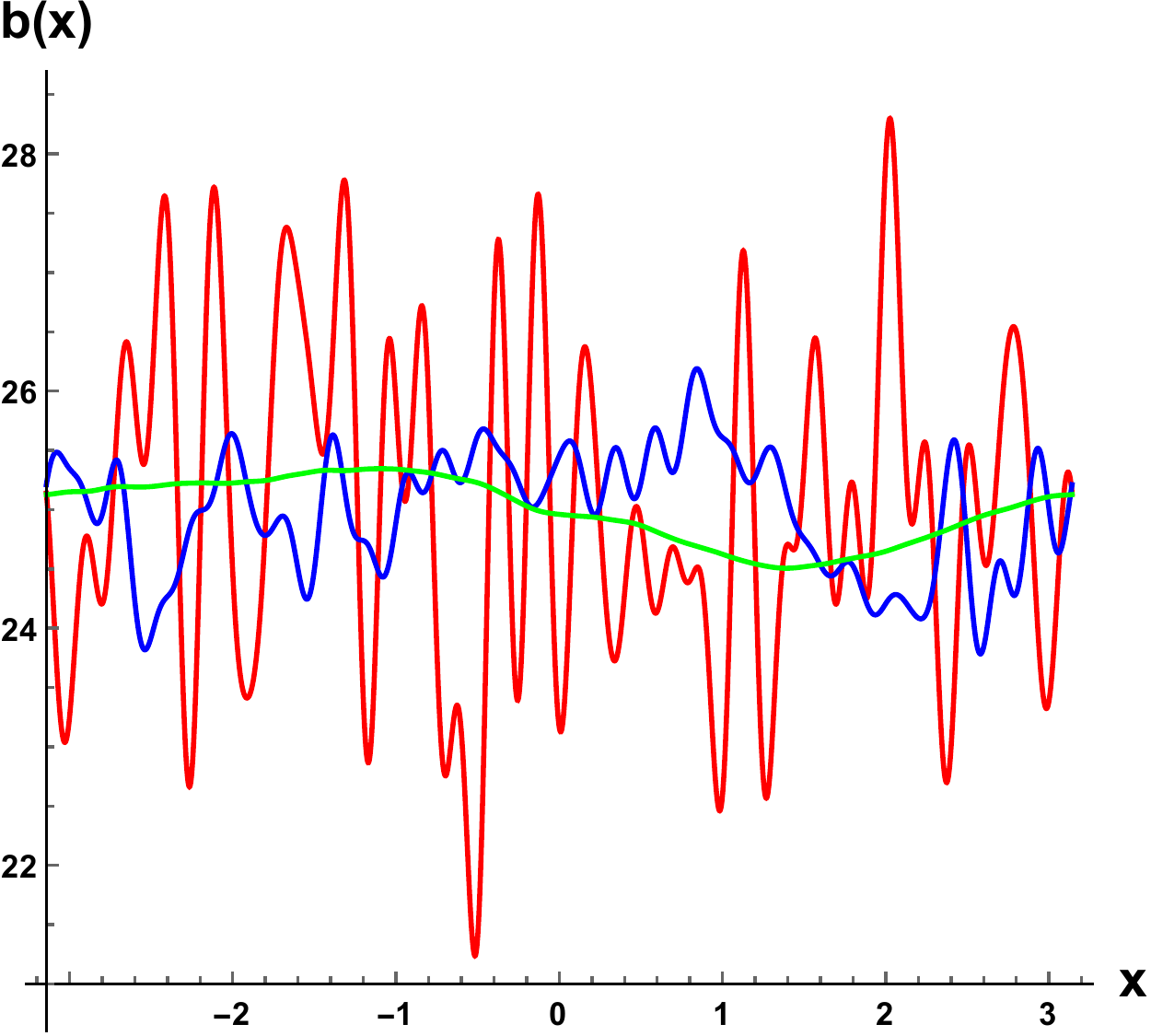}
\quad 
\includegraphics[width=0.45\textwidth]{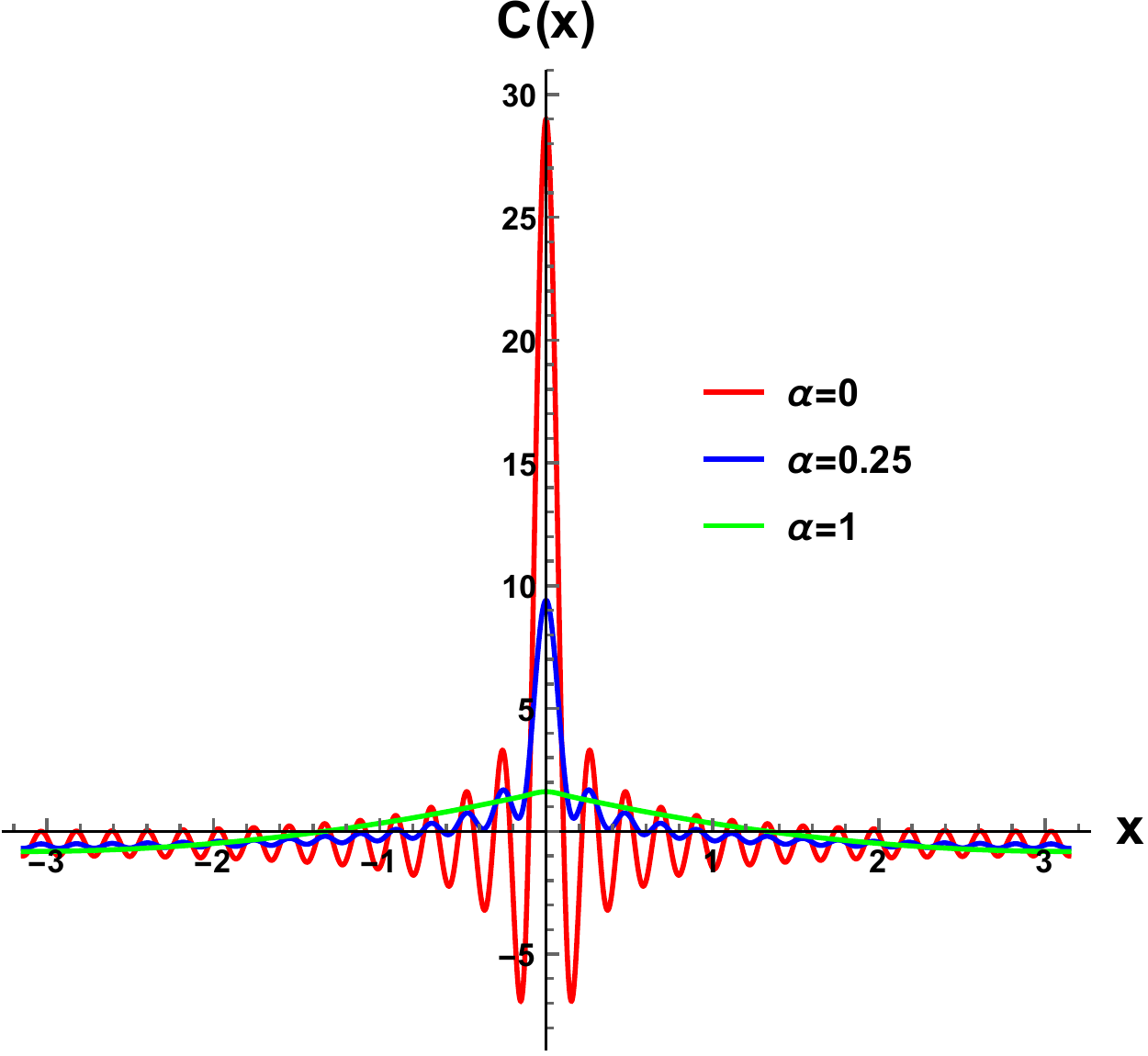}
  \caption{\textbf{Left: }Example of specific disorder realizations \eqref{eq:form} with $b_0=25$, $N=30$ and $\alpha=0,0.25,1$ (red, blue, green). \textbf{Right: }Corresponding correlation functions $\mathcal{C}(x)$.}
  \label{figex}
\end{center}
\end{figure}

In order to study the effects of one dimensional quenched disorder in the Weyl nodes distance we have introduced 1D Gaussian noise in the source of the gauge field $b(x)$ retaining the source for the scalar field $\phi$ constant $M(x)\equiv M$. Concretely we implement our disorder distribution (see fig.~\ref{figex}) using the spectral representation \cite{Shinozuka1991}:
\begin{equation}\label{eq:form}
    b(x) = b_0 + 2 \,\gamma \sum\limits_{i=1}^{N-1} \sqrt{S(k_i)} \,\sqrt{\Delta k} \, \cos\left( k_i \, x +\delta_i \right) \, .
\end{equation}
with equidistributed momenta $k_i=i\,\Delta k$ with separation $\Delta k=k_0/N$ and the random phases $\delta_i$ uniformly distributed in the interval $[0, 2\pi]$.
The parameter $\gamma$ measures the disorder strength. 
Since $x$ is a periodic coordinate with periodicity $L$, we may represent the disordered source using the spectral representation in terms of Fourier modes\footnote{In the plots, we have used $\tilde{x} = 2\pi\,x /L $ such that $\tilde{x}$ is $2\pi$-periodic.}; this amounts to choose $\Delta k=2\pi/L$. The quantity $k_{UV}\equiv k_0= 2\pi N/L$ provides a UV cutoff for our disorder. On the contrary, the quantity $k_{IR}\equiv k_0/N=2\pi/L$ sets the infrared scale. In order for our setup to be physically meaningful we need to satisfy the inequalities:
\begin{equation}\label{eq:inequalities}
k_{IR}\,\ll \,T\, \ll \,k_{UV}\,.
\end{equation}
Note that this can be achieved by a large enough size $L$ and a large enough number of modes $N$. The disorder average $\left\langle f \right\rangle_{dis}$  is defined as
\begin{equation}
 \left\langle f \right\rangle_{dis} = \int \prod\limits_{i=1}^{N-1} \frac{d \delta_i}{2\pi} f 
\end{equation}
and it is implemented numerically by averaging on a large number of different disorder realizations of the type \eqref{eq:form}.

We set the \textit{spectral density} $S$ of our signal to have the simple form $S(\xi)=\xi^{-2\alpha}$. The power $\alpha$ controls the correlation of our disordered distribution: positive and large $\alpha$ corresponds to positive and large correlation (see fig.~\ref{figex}). We put $\alpha=0$ where not explicitly stated otherwise.

The power $P$ of a signal is defined as as the autocorrelation function at $x=0$:
\begin{equation}\label{eq:power}
P\equiv\langle \Hat{b}(0)\hat{b}(0)\rangle=4\gamma^2\Delta k\sum_{i=1}^{N-1}\frac{1}{k_i^{2\alpha}}    \,=\,4\gamma^2\Delta k^{1\,-\,2\,\alpha}\,\mathcal{H}_{N-1}(2\,\alpha)\,, \end{equation}
where $\hat{b}(x) = b(x) - b_0$ and $\mathcal{H}_n$ is the n-th harmonic number. At finite correlation $\alpha \neq 0$ the power $P$ is a better suited measure of the disorder strength than the simple amplitude $\gamma$.

In the case of $\alpha=0,$ i.e. for $S(k_i)=1$, and in the limit $N \rightarrow \infty$, the distribution \eqref{eq:form} describes local, isotropic, Gaussian disorder:
\begin{equation}
\left\langle \hat{b}(x) \right\rangle_{dis} = 0 \, , \qquad\quad \left\langle \hat{b}(x) \,\hat{b}(y)   \right\rangle_{dis} = \gamma^2 \delta(x-y) \, ,
\end{equation}
with all the higher momenta vanishing.

The goal of this work is to investigate the fate of the quantum Weyl phase transition in presence of this type of one-dimensional Gaussian disorder. This implies that the dimensionality of our disorder is given by $[\gamma]=[b]-1/2$=1/2, where $b$ is the source that couples to the current operator dual to the bulk gauge field $A$. This is a relevant deformation and in this sense it is said not to satisfy Harris criterion. Since the relevant dimensionless quantity governing the phase transition is the ratio $b/M$ we expect similar results to hold for the case of an inhomogeneous source for the scalar field. \footnote{However, let us remark that this second option allows for the study marginal or irrelevant disorder, by appropriately setting the mass of the associated scalar field in the bulk. Another, numerically more challenging, option to achieve this is to increase the number of dimensions in which the fields are disordered.}

\section{Results: heating it up and making it dirty}\label{sec2}

In this section we discuss the effects of finite temperature and  quenched Gaussian disorder on the holographic Weyl semimetal topological quantum phase transition. 

\begin{figure}[htp]
\begin{center}
\includegraphics[width=0.48\textwidth]{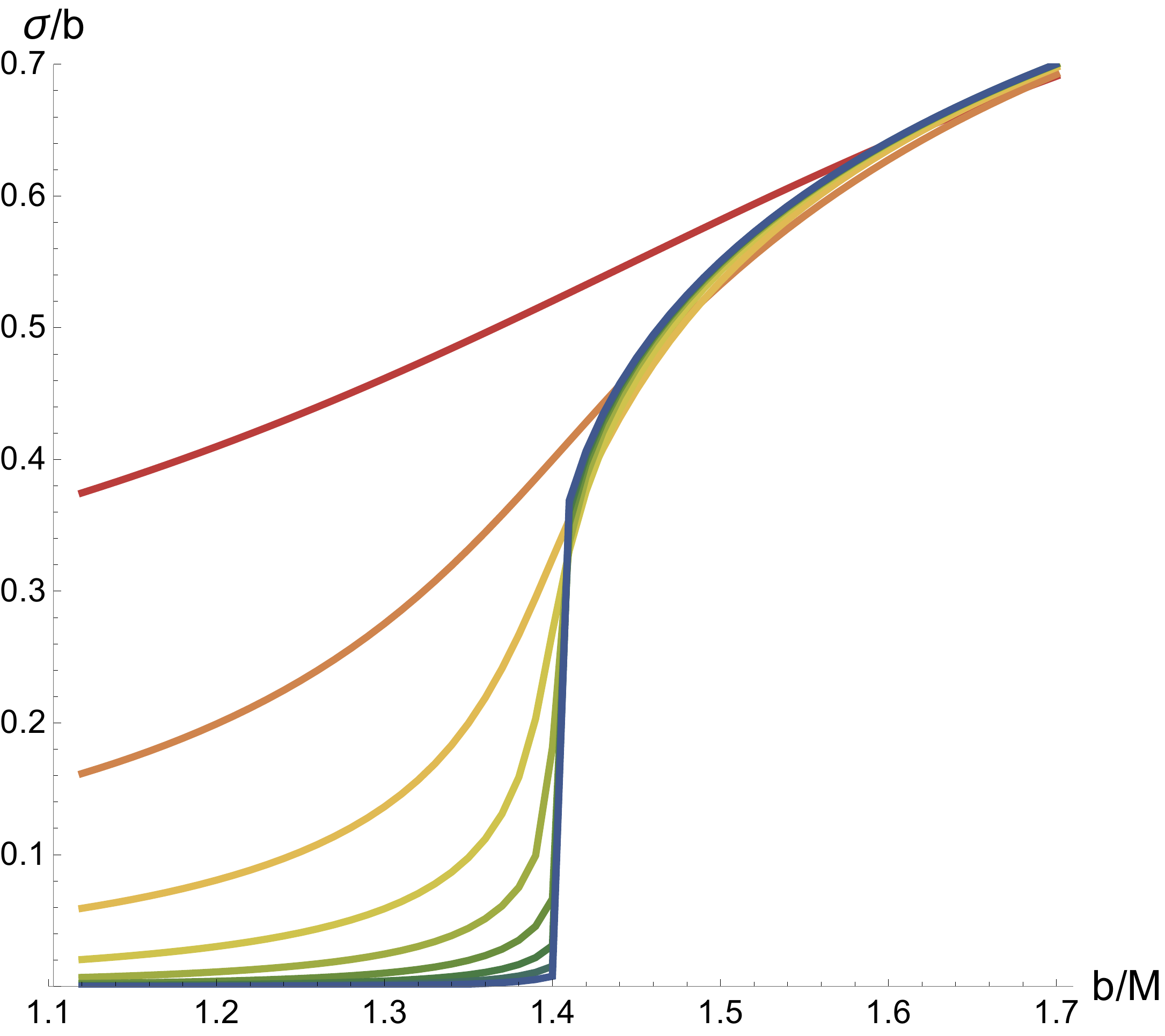}%
\quad
\includegraphics[width=0.48\textwidth]{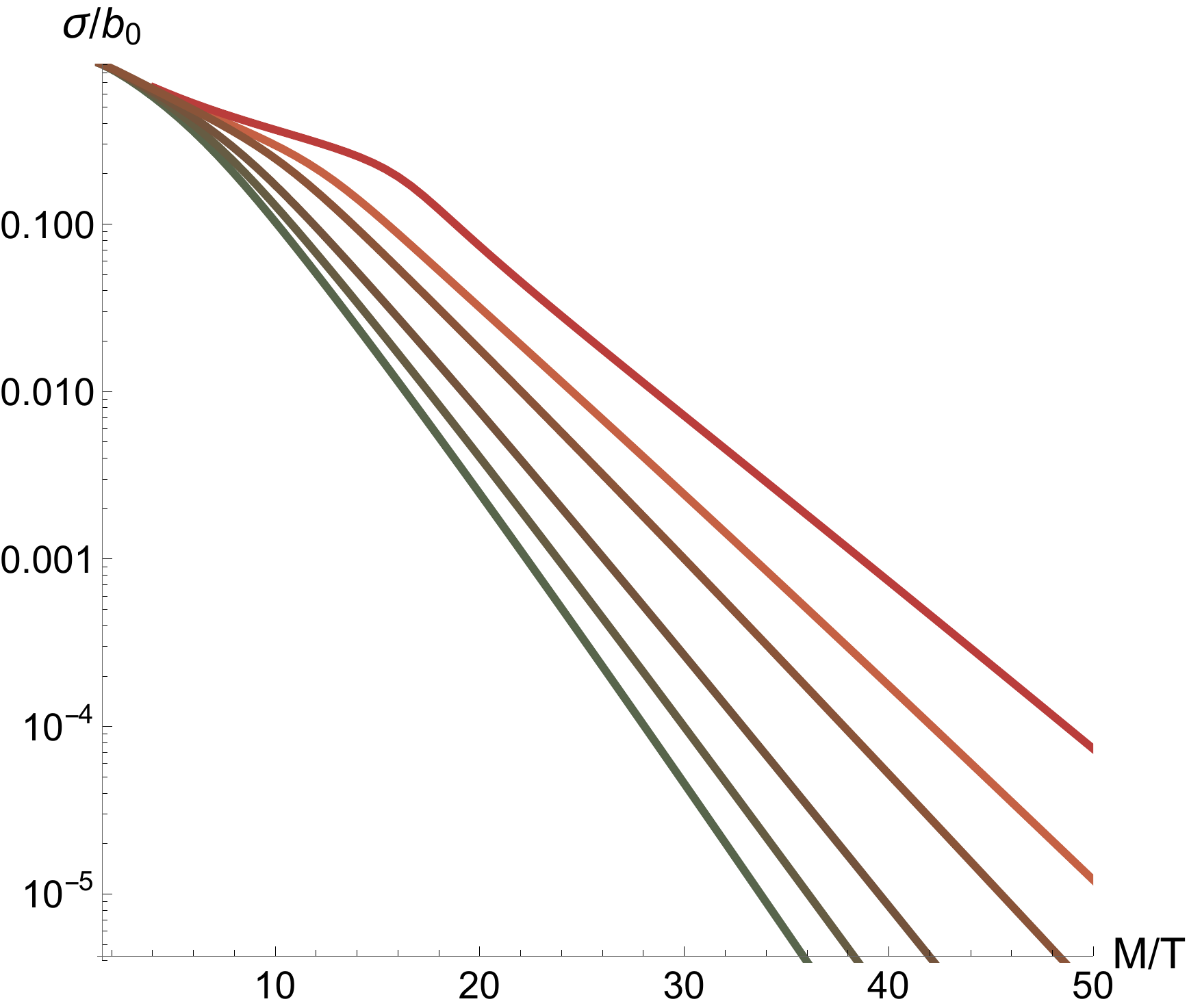}
  \caption{\textbf{Left: }AHC $\sigma$ against $\bar{b}$ for zero disorder and at finite temperature $T/M=1/\pi-1/(10\pi)$ (red-blue). At low enough temperatures (blue) a quantum phase transition occurs \label{fig:homogeneous}. \textbf{Right: } Semilogarithmic plot of the AHC $\sigma$ vs $M/T$ for fixed $\bar{b}={0.9-1.39}$ (green-red). At high enough M/T we fit the curves to a function of the type $e^{-c\,x^{\alpha}}$ where $x\equiv M/T$  finding $\alpha=1.05$. Close to the phase transition, at $\bar{b}=1.39$ (red line), a new scaling appears at low M/T. This is an effect of the quantum critical region (see fig.~\ref{figpower}). }
\end{center}
\end{figure}

\begin{figure}[htp]
\begin{center}
\includegraphics[width=0.48\textwidth]{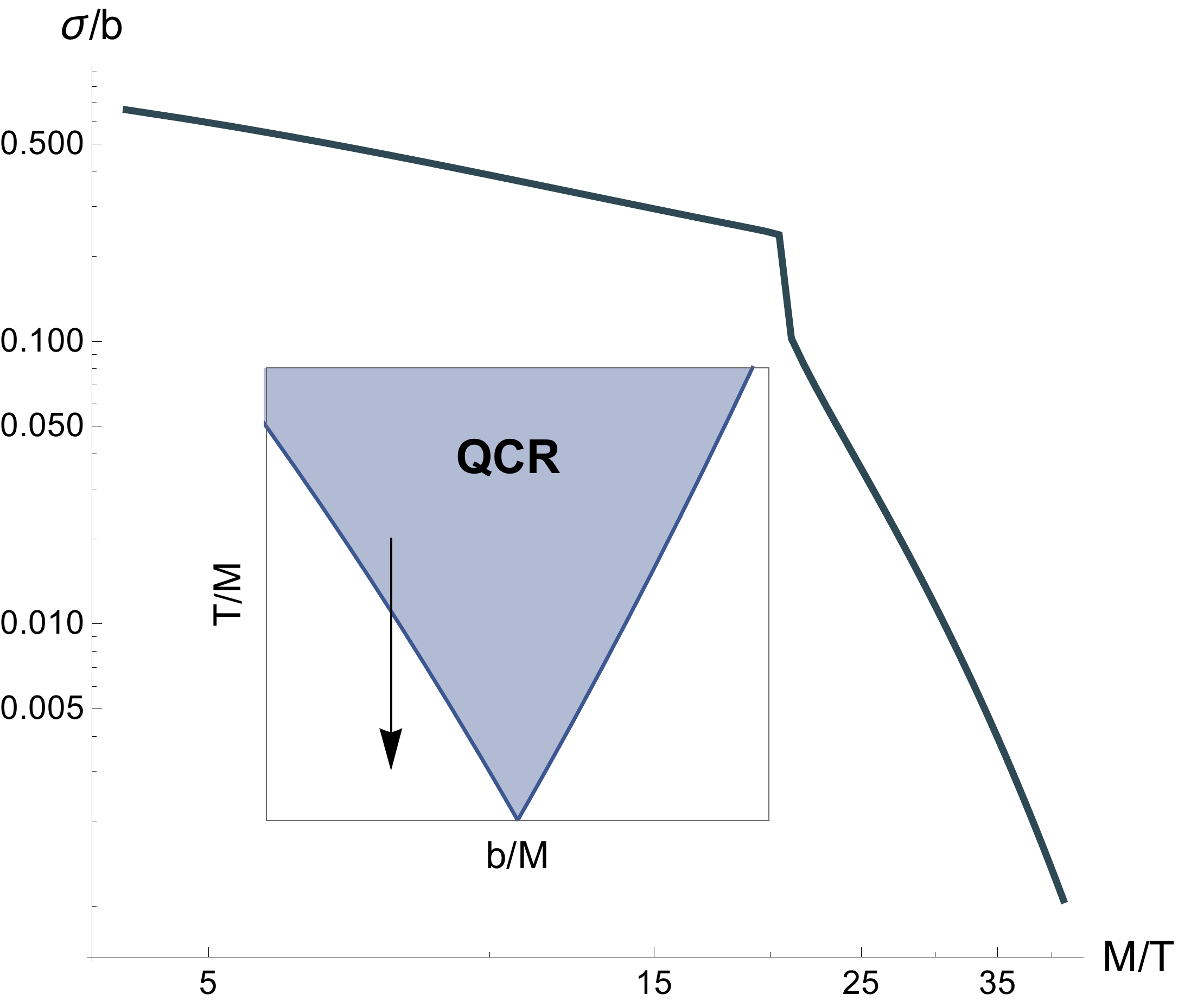}%
\quad
\includegraphics[width=0.48\textwidth]{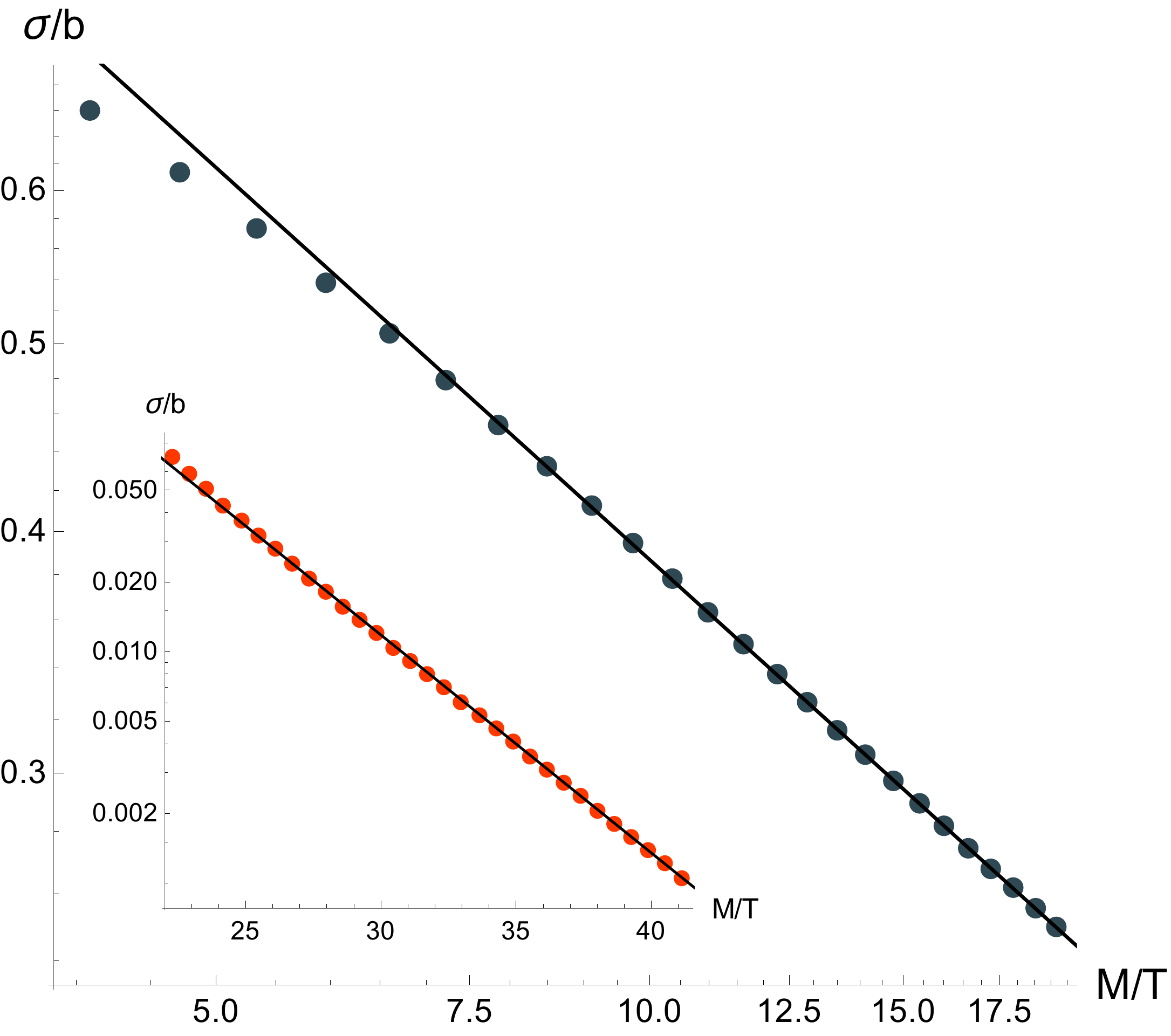}
  \caption{\textbf{Left: }Log-Log plot of the AHC $\sigma$ vs $M/T$ very close to the critical point $\bar{b}={1.4023}$. For small enough $M/T \gtrapprox 20$ the function fits very well a power law  $(M/T)^{-\nu}$ with $\nu=0.673$. At low enough temperature we abandon the critical region and the behaviour is no longer a power law as depicted in the inset. \textbf{Right:} Log-log zoom on the region $M/T \lessapprox 20$ shown in the left panel and fit (black line). In the inset we show a log plot for $M/T \gtrapprox 20$, where an exponential decay is recovered. Black line corresponds to the fit to $c_1e^{c_2\,M/T}$.  }
  \label{figpower}
\end{center}
\end{figure}

First, we study the effects of temperature on the clean QPT, which have been already partially analyzed in \cite{Landsteiner:2015pdh}. Finite temperature generically smoothens the sharp quantum phase transition into a crossover (see left panel in fig.~\ref{fig:homogeneous}). Eventually, for high enough temperature, the AHC is non-zero everywhere and the topologically trivial phase becomes inaccessible in the whole $\bar b$ range. We have analyzed the decay of the anomalous Hall conductivity $\sigma$ as a function of $T/M$ for several values of $\bar{b}$ moving towards the quantum critical point $\bar{b}_c$. Away from the quantum critical point, or in other words outside the \textit{quantum critical region}, we find an exponential decay of the form (see right panel in fig.~\ref{fig:homogeneous}):
\begin{equation}
\sigma\,\sim\,e^{-\,c\,M/T}\label{Teffects}
\end{equation}
where the parameter $c$ is not constant but rather it depends on $\bar{b}$ as shown in fig.~\ref{fig:homogeneous}. The exponential fall-off is consistent with the presence of a mass gap outside the quantum critical region which breaks the properties of scale invariance. Moving close to the quantum critical region, i.e. in the vicinity of $\bar{b}_c$, this behaviour is modified. As shown in fig.~\ref{figpower}, at big enough $T/M$ the decay follows a power law:
\begin{equation}
\sigma\,\sim\,\left(\frac{M}{T}\right)^{-\nu}\label{T2effects}
\end{equation}
were the ``critical exponent'' is found to be $\nu=0.673$. This is a clear signature of the presence of scale invariance inside the quantum critical region. 

Decreasing further the $T/M$ parameter we get back to an exponential decay. This can be understood from the inset in the left panel of fig.~\ref{figpower}: when getting close to zero temperature the critical region becomes thinner and thinner; at some point the system is again outside the quantum critical region and the exponential behaviour due to the mass gap is restored. Following the critical region up to $T/M=0$ requires a high numerical accuracy. We leave this for a future investigation including backreaction, which will allow us to directly probe the QCP at zero temperature.\\

\begin{figure}[htp]
\begin{center}
\includegraphics[width=0.45\textwidth]{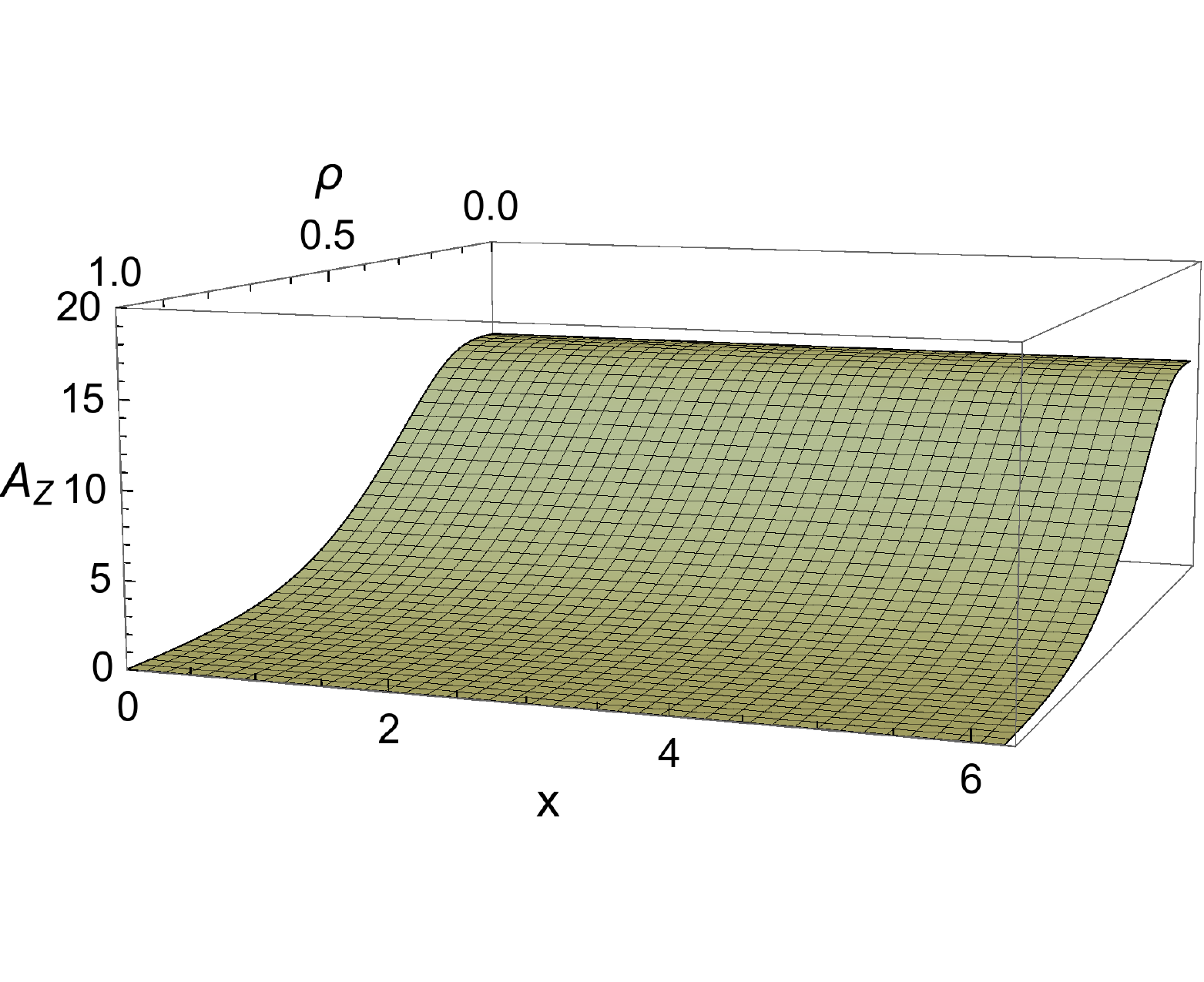}%
\qquad
\includegraphics[width=0.45\textwidth]{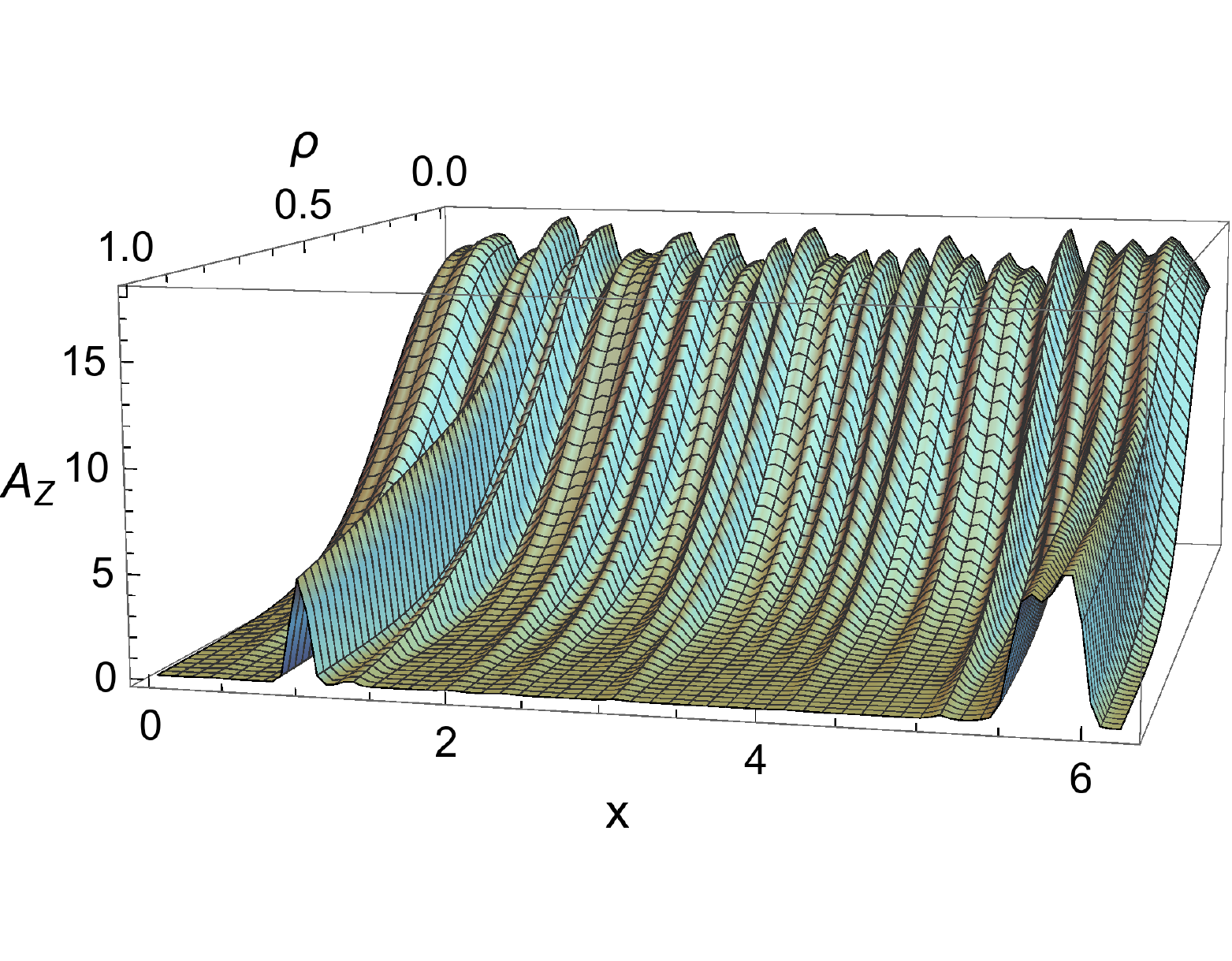}
  \caption{  \label{fig:configuration} A concrete configuration of the gauge field along the bulk for $T/M=1/(10\pi)$, $\bar{b}_0=1.35$ and $\bar{\gamma}=0,0.002$ (left, right).}
 \end{center}
\end{figure}

Now let us add some disorder into the system! We have computed the anomalous Hall conductivity in presence of Gaussian disorder \eqref{eq:form} in a wide range of temperatures $\bar{T}$ and $b_0/M\equiv\bar{b}_0\in[1-1.6]$ increasing the disorder strength\footnote{Let us remark that we have restricted $\gamma$ to values that do not generate negative Weyl node distances .} $\gamma$. To do this consistently we define another dimensionless ratio, $\bar{\gamma} \equiv \gamma\,\sqrt{\Delta k}/M$, which from now on we refer to simply as the disorder strength.

In order to fulfill the requirement \eqref{eq:inequalities} we have chosen $L=20\pi$ and $N=30$. We explicitly checked that the qualitative results are independent of the cutoffs. A large N scaling analysis would be necessary to make further and robust statements about the concrete quantitative results, like the critical exponents values. In order to have enough resolution to render grid size independent results we used grids of 51x101 (z,x directions respectively). Moreover compute $\sim 50$ different random realizations for each point in parameter space in order to obtain a small enough variance of the mean of the AHC. See fig.~\ref{figex} for a concrete realization of random source for the gauge field $A_z$. 

\begin{figure}[htp]
\begin{center}
\includegraphics[width=0.45\textwidth]{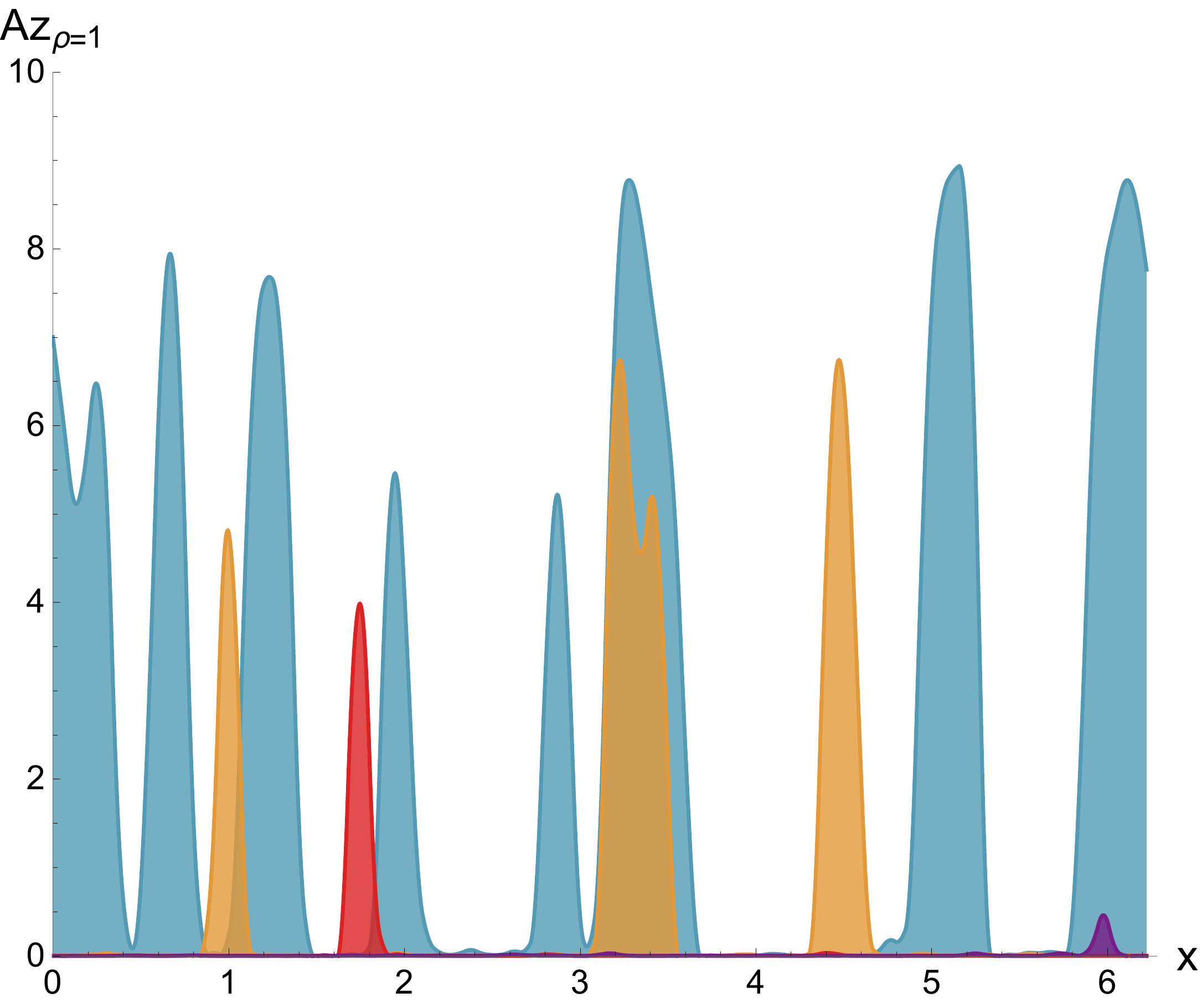}
\qquad
\includegraphics[width=0.45\textwidth]{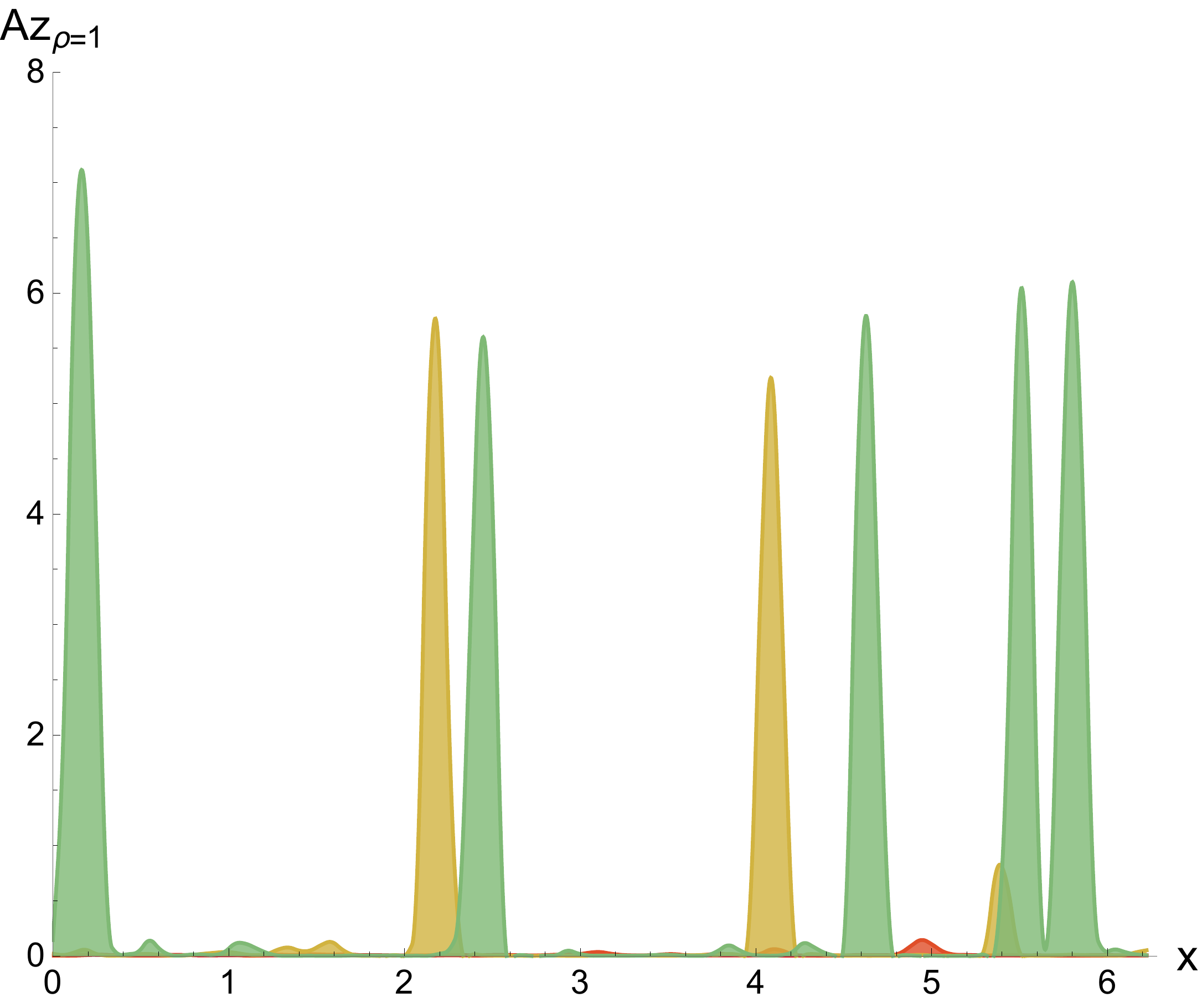}%
  \caption{  \label{fig:rarerare} The profile of the gauge field $A_z$ at the horizon and the appearance of the ``rare'' regions. \textbf{Left: }At fixed disorder strength $\bar\gamma=0.002$ approaching the critical point. We take $\bar{b}_0=1.25,\,1.3,\,1.35,\,1.4$  (purple, red, orange, blue). \textbf{Right :} At fixed $\bar{b}_0=1.35$ and increasing strength  $\bar\gamma=0.01,\, 0.02,\, 0.025$ (red, yellow, green).}
 \end{center}
\end{figure}%

In fig.~\ref{fig:configuration} we show the bulk profile of the gauge field $A_z(\rho)$ at $M/T=10\pi$ and $\bar{b}_0=1.35$ for a concrete realization of the random disorder described previously. We compare the results with the clean and homogeneous system. In absence of disorder (left panel of fig.~\ref{fig:configuration}) the gauge field vanishes everywhere along the horizon, giving rise to a zero anomalous Hall conductivity. On the other hand, as shown in the right panel of fig.~\ref{fig:configuration}, the presence of a disordered source triggers the appearance of localized areas at the horizon where the gauge field does not vanish. As a consequence the integrated AHC acquires a finite value which was not present in the clean system. 
   
In fig.~\ref{fig:rarerare} we show several realizations of the profile of the gauge field at the horizon, as a function of the spatial direction $x$. In the left panel of the figure we show the profile for fixed disorder strength $\bar\gamma$ and increasing $\bar{b}_0$. The appearance of localized areas which have locally undergone the phase transition is apparent. As the system is tuned closer the to the critical $\bar{b}_{0}$ these ``rare'' regions become broader and less rare. Similar behavior is found for fixed $\bar{b}_0$ and increasing disorder strength $\bar\gamma$, as shown in the right panel of fig.~\ref{fig:rarerare}. It is tempting and suggested by our results to interpret these areas as ``rare'' regions discussed in the condensed matter literature \cite{2006JPhA...39R.143V}. \\

\begin{figure}[htp]
\begin{center}
\includegraphics[width=0.46\textwidth]{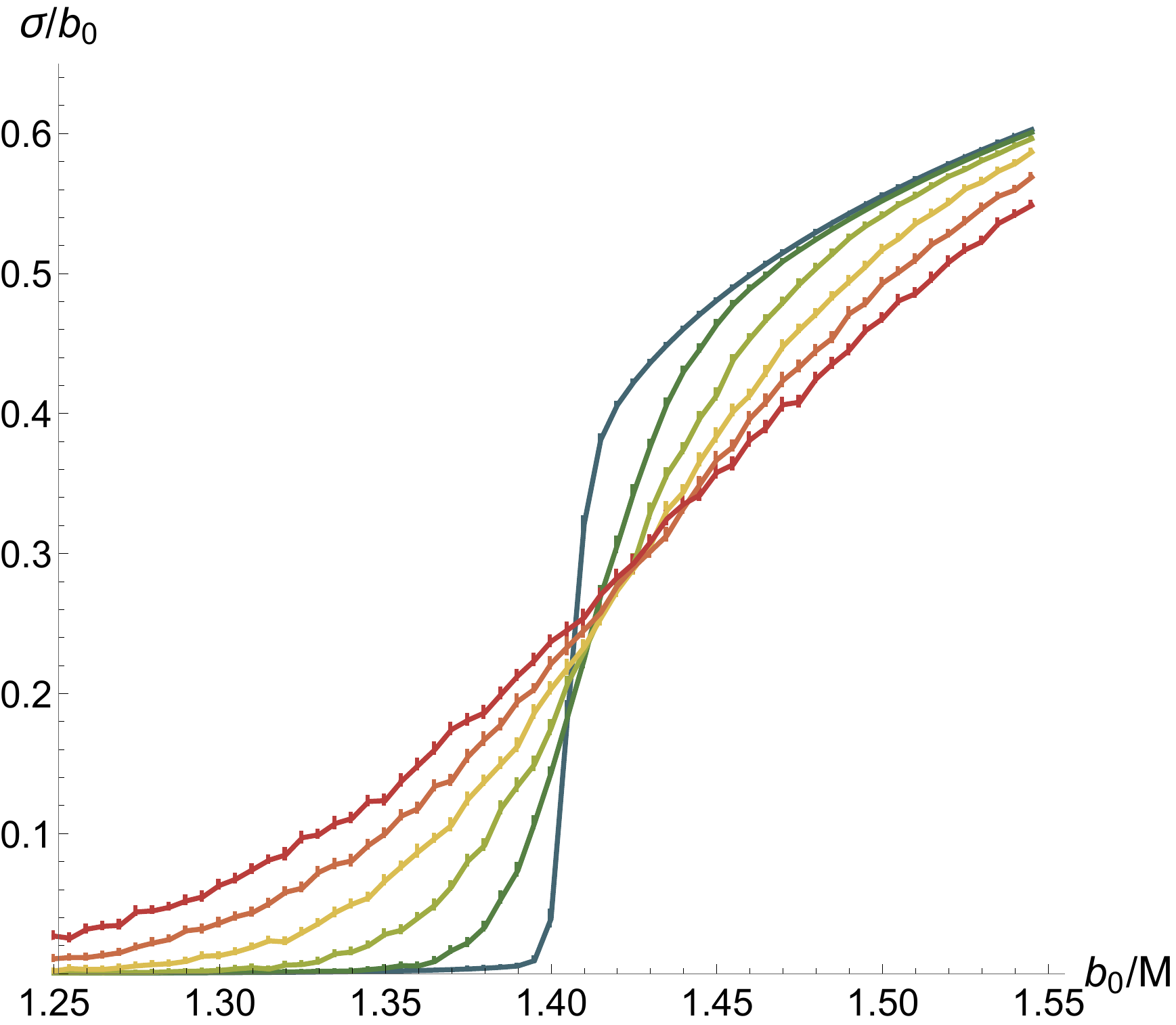}
\includegraphics[width=0.47\textwidth]{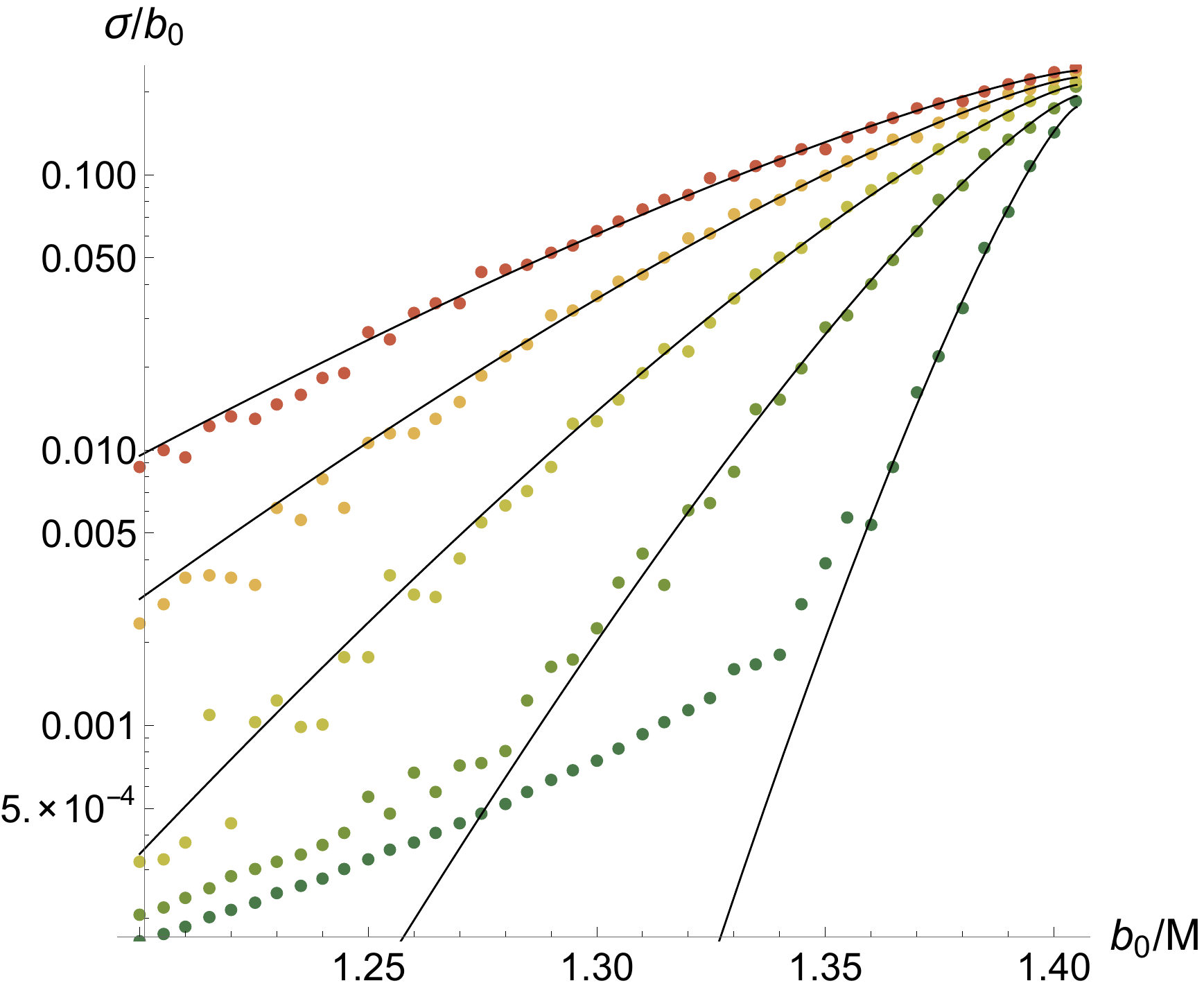}
  \caption{\textbf{Left:} Anomalous Hall conductivity as a function of  vs $\bar{b}_0$ for $\bar{\gamma}\in[0.003,0.035]$ (blue-red). \textbf{Right:} Semilogartihmic plot of the AHC against $\bar{b}_0$ for $\bar{\gamma}\in[0.001,0.035]$ (green-red). Black lines correspond to the exponential fit \eqref{diseffects}.}
   \label{fig:figdis}
\end{center}
\end{figure}
\begin{figure}[htp]
\begin{center}
\includegraphics[width=0.47\textwidth]{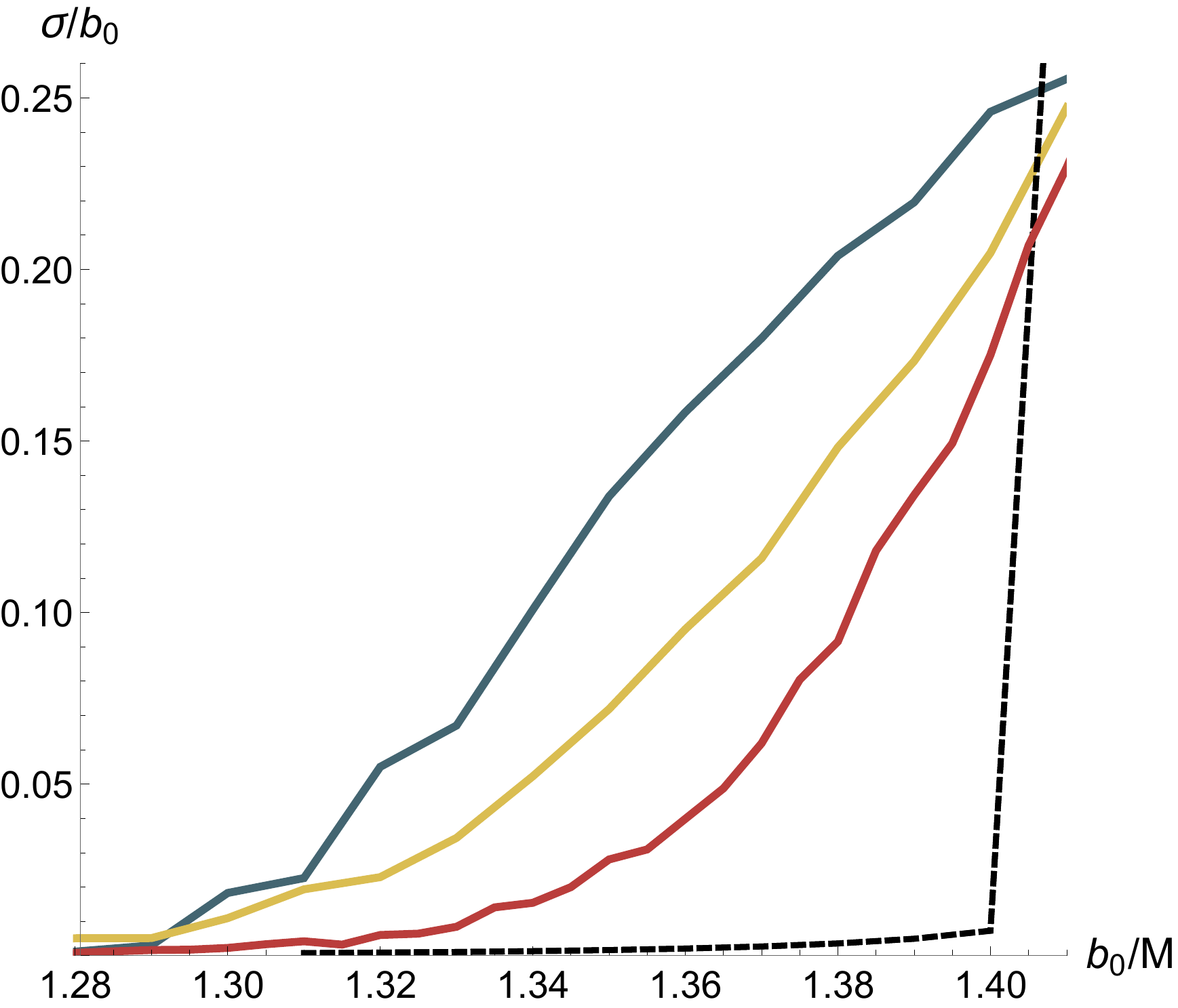}
     \quad   \quad 
\includegraphics[width=0.47\textwidth]{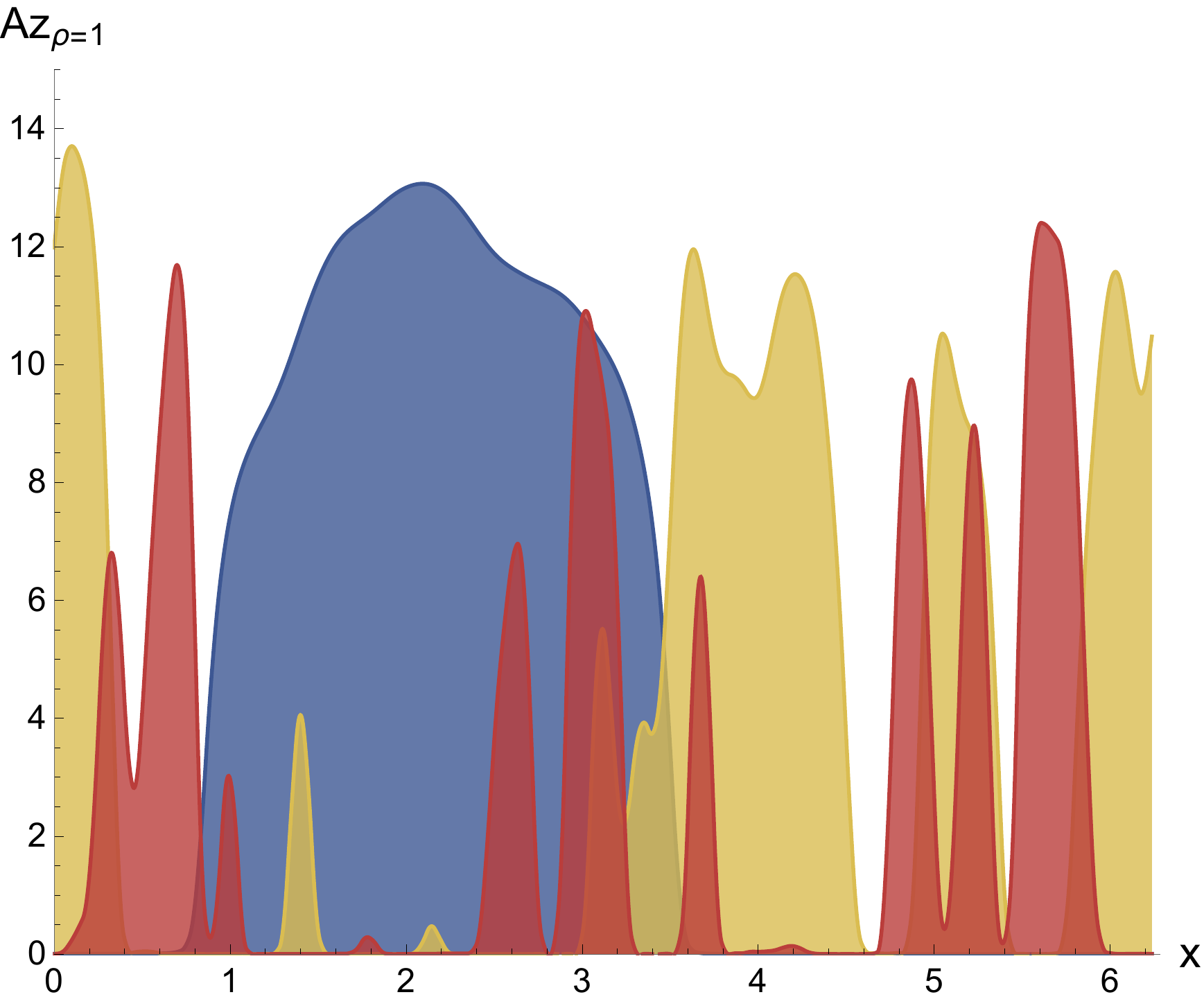}
     \caption{\label{fig:correlation} \textbf{Left:} AHC close to the critical $\bar{b}_0$ for fixed $P=0.15$ and $M/T=10/\pi$ and increasing correlation $\alpha=0,\,0.5\,,2$ (red, yellow, blue). Black dashed line shows the AHC at zero disorder. \textbf{Right:}Evolution of the rare regions with alpha for $\alpha=0,\,0.5,\,2$ (red, yellow, blue) at fixed $P=0.15$. Localized regions become broader, as expected for increasingly correlated disorder.
     }
 \end{center}
\end{figure}

We now turn to the study of the quantum phase transition at finite disorder. In fig.~\ref{fig:figdis} we show the anomalous Hall conductivity at low temperature $T/M=1/(10\pi)$ as a function of $\bar{b}_0$ and for increasing disorder strength $\bar\gamma$. The sharp QPT gets smeared out by the presence of disorder as a direct consequence of the presence of the localized regions. Close to the citical value $\bar{b}_{0}\sim1.4023.$ the smearing is well approximated by an exponential tail of the form:
\begin{equation}
  \sigma_{xy} \sim\,c_1\,e^{c_2(1.4023 -\bar{b}_0)^{c_3}}\label{diseffects}\,,
\end{equation}
as depicted in the right panel of fig.~\ref{fig:figdis}. We have found numerically the parameter $ c_3$ to be independent of the disorder strength and in all cases compatible with the value $c_3\sim1.28$. On the contrary, the other coefficients $c_1,\,c_2$ appear to be strongly dependent on the disorder strength $\bar{\gamma}$.
This functional dependence, as well as the independence of the power $c_3$ from the details of the randomness, is in agreement with the results obtained with Optimal Fluctuation Theory for composition tuned quantum smeared phase transitions, where $c_3=2-d/\phi$, with $\phi$ the so-called \textit{finite size shift exponent}, \textit{i.e.} just dependent on the details of the clean QPT (see for example \cite{2011PhRvB..83v4402H}).\\

Additionally we can study the role of the disorder correlation on the quantum phase transition and its smearing. In order to do that we fix the power $P$ \eqref{eq:power} of our disorder signal and the number of modes $N$ and we consider several realizations with different $\alpha$, ranging from the  uncorrelated case $\alpha=0$ to highly correlated disorder, \textit{i.e.}  $\alpha=2$ (see fig.~\ref{figex}). 
 As shown in fig.~\ref{fig:correlation} we find that the disorder correlation plays indeed a role on the QPT. Concretely we find that positive correlation increases the smearing of the order parameter, in agreement with \cite{PhysRevLett.108.185701,2013AIPC.1550..263N,0295-5075-97-2-20007}. On the right panel of fig.~\ref{fig:correlation} we show a concrete realization of the gauge field at the horizon for fixed $P$ and increasing correlation length. We see indeed how increasing correlation gives rise to broader, less rare, regions that have undergone the phase transition.

\begin{figure}[htp]
\begin{center}

     \includegraphics[width=0.47\textwidth]{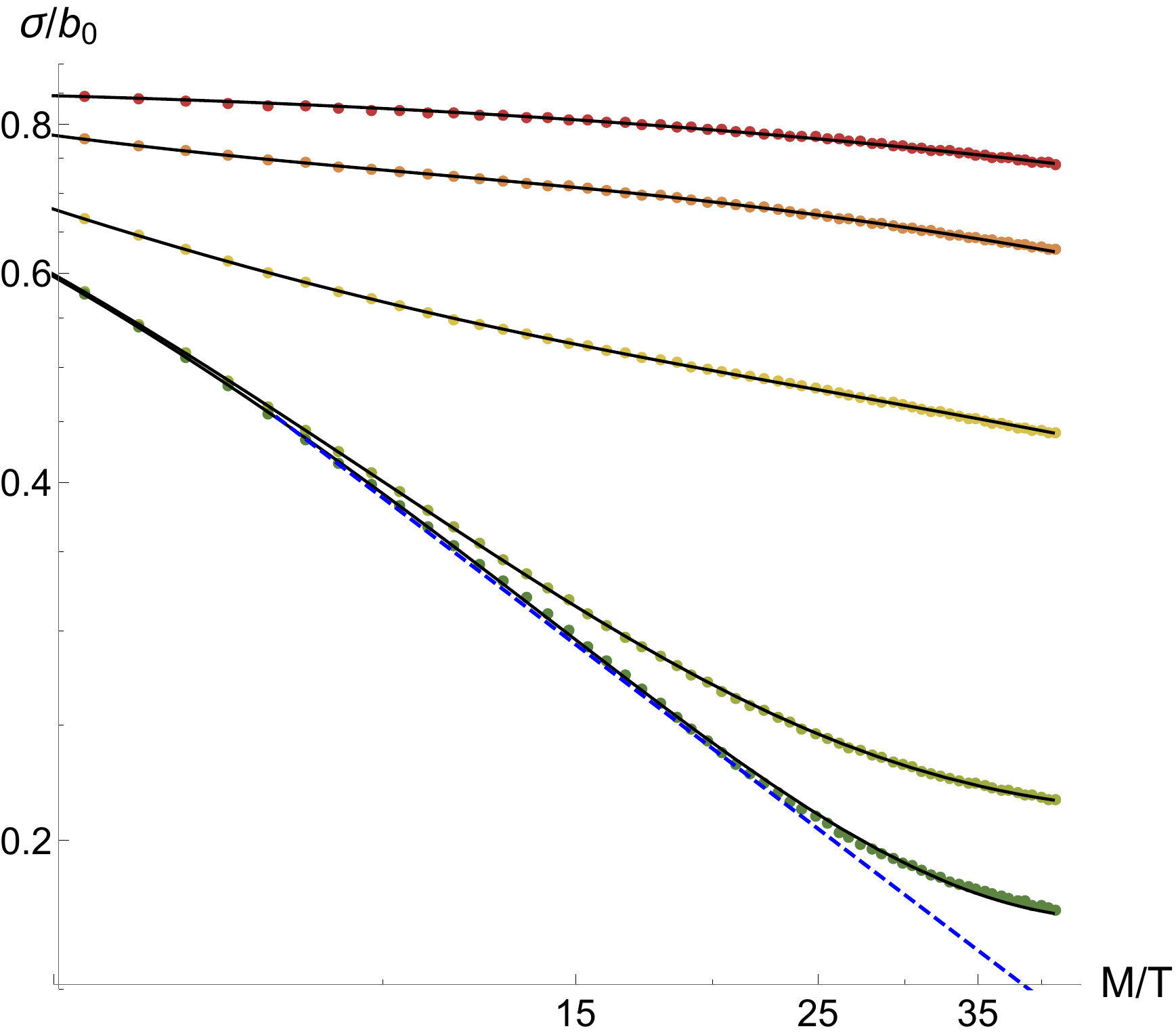}
      \includegraphics[width=0.48\textwidth]{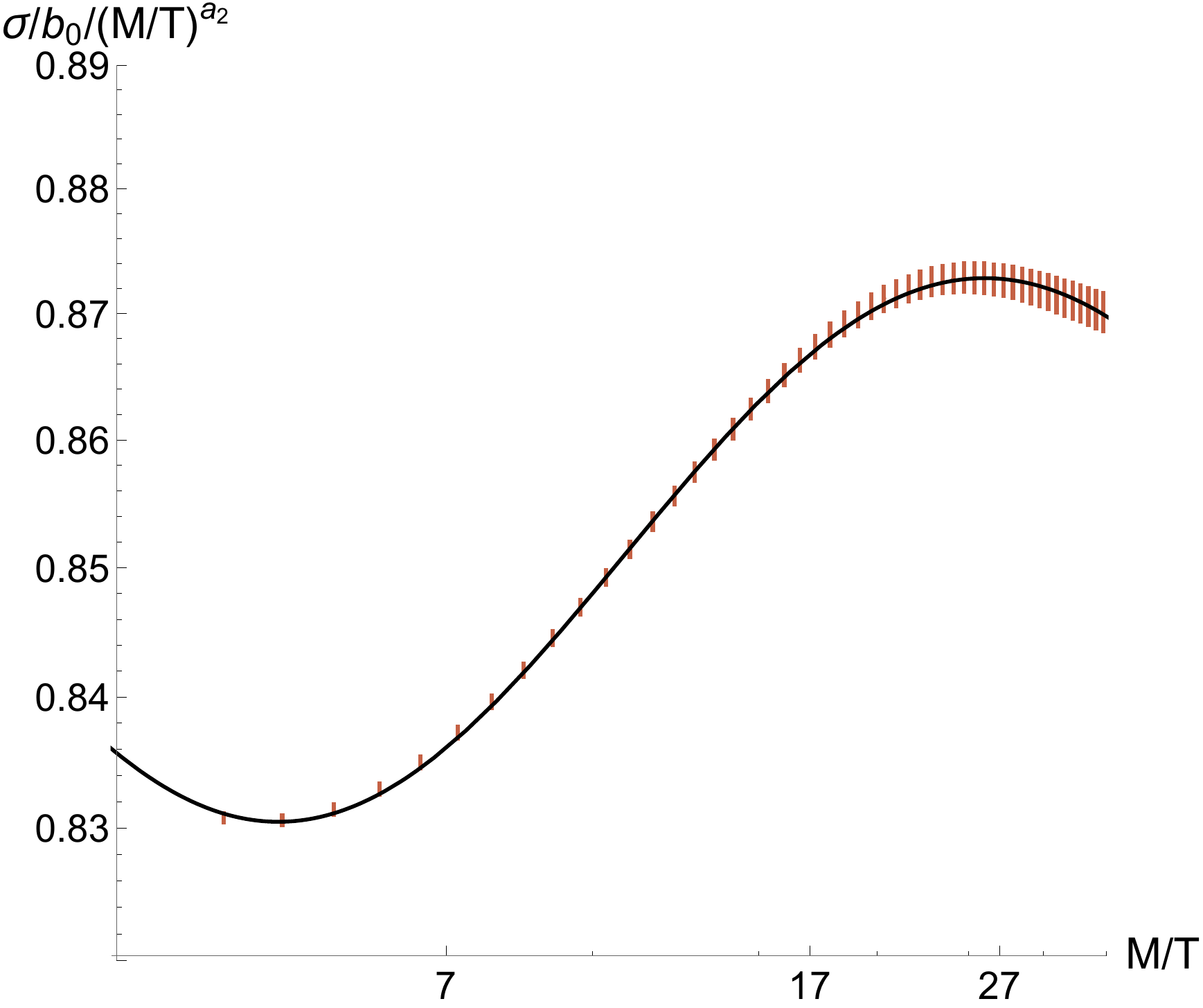}
     \caption{  \label{Tdowncritical}\textbf{Left:} Log-Log plot at the critical point $\bar{b}_0=1.4023$ with increasing disorder strength $\bar\gamma=(0.003-0.6)$ (green-red). For high enough disorder the data fit well to a log-oscillatory function of the type: $t(x)=a_1\, x^{a_2}(1+a_3 \sin{[a_4 \log (x)+a_5])}$ (black lines). At low disorder (green line) the data approximates the power law decay found at zero disorder and shown in blue-dashed line. \textbf{Right:} Data at $\bar\gamma=0.4$  (orange line in the left panel), divided by $(M/T)^{a_2}$ with exponent $a_2$ given by the fit. Oscillations become apparent.}
 \end{center}
\end{figure}

Finally we study the fate of the finite temperature scaling at finite disorder. We have found that the power law decay in the critical region is modified in a very interesting way. In particular, we notice a ``wiggling'' of the decay which is consistent with a log-oscillatory form (see fig.~\ref{Tdowncritical}). To be precise, we find a behaviour consistent with the form:
\begin{equation}
\sigma\left(M/T\right)\,\sim\,a_1\,\left(M/T\right)^{a_2} (1+a_3 \sin{[a_4 \log (M/T)+a_5])}
\end{equation}
inside the quantum critical region.

This behaviour is known to be connected  with the appearance of discrete scale invariance at the disordered fixed point \cite{SORNETTE1998239}\footnote{See \cite{Flory:2017mal} for some recent discussions about discrete scale invariance in the context of holography.} and has been already observed in holography in \cite{Hartnoll:2015rza}. Unfortunately at very low temperature and high disorder our probe approximation and the numerical methods we used are not reliable anymore. Following our preliminary indications it would be very interesting to investigate further this point including full backreaction.

\section{Conclusions}\label{sec3}
In this paper we have studied the effects of temperature and quenched disorder on a quantum phase transition in holography. In particular we focused on the probe limit analysis of the holographic Weyl semimetal quantum phase transition (QPT) \cite{Landsteiner:2015lsa,Landsteiner:2015pdh} in presence of 1-D Gaussian disorder.

First we have investigated the effects of temperature on the clean QPT appearing at $\bar{b}_{c}\approx 1.4023$. Finite temperature enhances the anomalous Hall conductivity (AHC) and tends to destroy the trivial phase where the AHC vanishes. The fall-off of the AHC towards $T\rightarrow 0$ appears to be:
\begin{itemize}
    \item exponential $\sim e^{-\Delta /T}$ for values of the external parameter far from the critical point $\bar{b}<\bar{b}_c$ or more precisely outside the \textit{quantum critical region} (see sketch in fig.~\ref{fig1}). This is consistent with the presence of a finite mass gap $\Delta$.
    \item power law $\sim T^{-\nu}$ around the critical point $\bar{b}\approx \bar{b}_c$, \textit{i.e.} inside the \textit{quantum critical region}. This is just a manifestation of the presence of scale invariance at criticality.
\end{itemize}
Moreover, we have introduced disorder into the system and studied the fate of the QPT. Our results show that the presence of disorder induces the appearance of localized \textit{rare} regions in the spatial profile of the vector field at the horizon (see fig.~\ref{fig:configuration} and \ref{fig:rarerare}) which account for a finite integrated AHC $\tilde{\sigma}$, absent at zero randomness. The smearing of the sharp QPT (see fig.~\ref{fig:figdis}) is the main consequence of the presence of disorder and it appears to be consistent in its functional form with the expectations from CM theory \cite{2006JPhA...39R.143V} and optimal fluctuations arguments \cite{2011PhRvB..83v4402H}.

In addition, modifying the correlation of our disorder distribution, we proved that correlation indeed plays a role in the smearing of the quantum phase transition as expected from condensed matter theory studies \cite{PhysRevLett.108.185701,2013AIPC.1550..263N,0295-5075-97-2-20007}. Our findings, in qualitative agreement with the results therein, show that positive correlation enhances the order parameter at the tail and therefore the smearing of the QPT (see fig.~\ref{fig:correlation}).

Finally we have investigated the temperature fall-off of the integrated AHC $\tilde{\sigma}$ in presence of disorder. We notice that at strong enough disorder and within the quantum critical region the AHC exhibits log-oscillatory behaviours as a function of T which seem consistent with the emergence of a disordered fixed point enjoying discrete scale invariance \cite{SORNETTE1998239}.

The results of the present paper open several questions and future directions. The main, numerical demanding, but very valuable direction, would be to study the same setup with full backreaction. That would allow for several improvements: full control on the bulk solution up to zero temperature, complete analysis of the disordered fixed points and possibility of computing more observables like entropy, heat capacity and longitudinal conductivities and viscosities (\cite{Landsteiner:2015pdh,Landsteiner:2016stv}) in presence of disorder. On the same ground it would be extremely interesting to analyze in more details the presence of the log-oscillatory structures.
It could also be possible to apply the same techniques we used to introduce disorder into other holographic QPTs, for example \cite{DHoker:2010onp,DHoker:2012rlj}, to test to which extent our results are universal. 

We hope to come back to some of these questions in the near future.

\section*{Acknowledgements}
We thank Panagiotis Betzios, Danny Brattan, Carlos Hoyos and Tomas Vojta for useful discussions and comments about this work and the topics considered. We are specially grateful to Daniel Arean, Sean Hartnoll and Karl Landsteiner for reading a preliminary version of the manuscript and providing several insightful comments and suggestions. We are grateful to Andre Sternbeck for his support with the local cluster.\\
MB is supported in part by the Advanced ERC grant SM-grav, No 669288. AJ and SM acknowledges financial support by Deutsche Forschungsgemeinschaft (DFG) GRK 1523/2.\\
MB would like to thank the Adolfo Iba\~nez University of Vi\~na del Mar (Chile), the organizers of the workshop ``Holography and Supergravity 2018'', the ICTP South American Institute for Fundamental Research (ICTP-SAIFR) and the University of Sao Paulo (USP) for the warm hospitality during the completion of this work. MB would like to thank Marianna Siouti for the unconditional support \panda.


\appendix

\section{Anomalous Hall conductivity}\label{app2}
In this appendix we provide details about the computation of the anomalous Hall conductivity in the inhomogeneous holographic Weyl semimetal. We consider an electric field in the $x$-direction and define the Hall conductivity as the response in the transverse $y$ direction as:
\begin{equation}
\left\langle j_y \right\rangle = \sigma_{yx} \,E_x \, .
\end{equation}
In order to determine $\sigma$ we consider the following fluctuations:
\begin{equation} 
\delta V = (\tilde{v}_t(x,\rho) \, e^{- i \omega t}, \tilde{v}_x(x,\rho) \, e^{- i \omega t}, \tilde{v}_y(x,\rho) \, e^{- i \omega t},0,0) \, .
\end{equation} 
We impose in-going boundary conditions at the horizon as follows: 
\begin{eqnarray}
\tilde{v}_t(x,\rho) & = & v_t(x,\rho) \, (1-\rho)^{1- i \omega/4} \, ,\\
\tilde{v}_x(x,\rho) & = & v_x(x,\rho) \, (1-\rho)^{- i \omega/4} \, ,\\
\tilde{v}_y(x,\rho) & = & v_y(x,\rho) \, (1-\rho)^{- i \omega/4} \, , 
\end{eqnarray}
where $v_\mu(x,\rho)$ are regular functions at the horizon.

We are interested in the DC Hall conductivity. Hence we have to solve the equations of motion for the fluctuations only to first order in $\omega.$ Moreover, we also perform an expansion in $\kappa$ and neglect terms of order $\kappa^2$
\begin{eqnarray}
v_t(x,\rho) & = & v_t^{(0,0)}(x,\rho) + v_t^{(1,0)}(x,\rho) \, \omega  + v_t^{(0,1)}(x,\rho) \, \kappa + v_t^{(1,1)}(x,\rho) \, \omega\, \kappa +  \dots \, , \\
v_x(x,\rho) & = & v_x^{(0,0)}(x,\rho) + v_x^{(1,0)}(x,\rho) \, \omega  + v_x^{(0,1)}(x,\rho) \, \kappa + v_x^{(1,1)}(x,\rho) \, \omega \,\kappa + \dots \, ,  \\
v_y(x,\rho) & = & v_y^{(0,0)}(x,\rho) + v_y^{(1,0)}(x,\rho) \, \omega  + v_y^{(0,1)}(x,\rho) \, \kappa + v_y^{(1,1)}(x,\rho) \, \omega \,\kappa + \dots \, ,
\end{eqnarray}
where the neglected terms (indicated by dots) are of order $\mathcal{O}(\kappa^2, \omega^2,\dots)$. In the following we discuss now the differential equations for $v_t^{(m,n)}, v_x^{(m,n)}$ and $v_y^{(m,n)}$. We obtain the following equations of motion for $v_t^{(0,0)}, v_x^{(0,0)}$ and $v_y^{(0,0)}$ :
\begin{eqnarray}\nonumber
(1+\rho+\rho^2+\rho^3) \, v_t^{(0,0)} - (1+\rho)^2 (1+\rho^2) \, \partial_\rho v_t^{(0,0)} & & \\ 
+\rho \, (1-\rho^4) \, \partial^2_\rho v_t^{(0,0)} + \rho \, 
\partial^2_x v_t^{(0,0)} &=& 0 \, , \\
(1+3\rho^4) \, \partial_\rho v_x^{(0,0)} - \rho \, (1-\rho^4) \partial^2_\rho v_x^{(0,0)} &=& 0 \, , \\
(1+3\rho^4) \, \partial_\rho v_y^{(0,0)} - \rho \, (1-\rho^4) \partial^2_\rho v_y^{(0,0)} -\rho \, \partial^2_x v_y^{(0,0)}  &=& 0 \, ,
\end{eqnarray}
while the constraint evaluated at the horizon reduces to:
\begin{equation}
v_t^{(0,0)}(x,\rho=1) - \partial_x v_x^{(0,0)}(x,\rho=1) = 0 \, .
\end{equation}
Let us first consider the order $\kappa^0 \omega^0$. Switching on only an electric field in $x$-direction we impose the following boundary conditions at the AdS boundary $\rho=0$,
\begin{equation}
v_t^{(0,0)} (x,0) = 0 \, ,\qquad v_x^{(0,0)}(x,0)=1 \, , \qquad v_y^{(0,0)}(x,0) = 0 \, .
\end{equation}
The solution respecting these boundary conditions reads:
\begin{equation}
v_t^{(0,0)} (x,\rho) = 0 \, ,\qquad v_x^{(0,0)}(x,\rho)=1 \, , \qquad v_y^{(0,0)}(x,\rho) = 0 \, .
\end{equation}
Next, we proceed with order $\omega^1 \kappa^0.$ The functions $v_t^{(1,0)}$ and $v_y^{(1,0)}$ are trivial:
\begin{equation}
v_t^{(1,0)}(x,\rho)=0 \, , \qquad v_y^{(1,0)}(x,\rho)=0 \, .
\end{equation}
Note that the differential equation for $v_x^{(1,0)}$ is sourced by $v_x^{(0,0)}$, however, the constraint does not impose a non-trivial $x$-dependence. Moreover, the functions appearing in the differential equation are not $x$-dependent. Hence, the solution reads  
\begin{equation}
v_x^{(1,0)}(x,\rho)= -\frac{i}{4}  \left( \log(1+\rho) - \log(1+\rho^2) \right) \, ,
\end{equation}
where we imposed that $v_x^{(1,0)}(x,0)=0$ as well as regularity of $v_x^{(1,0)}$ at the horizon.

The order $\omega^0 \kappa^1$ is trivial since the electric field is perpendicular to the axial magnetic field induced by $A_z(x).$ Hence the solution (which also satisfies the constraint) reads
\begin{equation}
v_t^{(0,1)}(x,\rho)=0 \, , \qquad v_x^{(0,1)}(x,\rho)=0 \, , \qquad v_y^{(0,1)}(x,\rho)=0 \, .
\end{equation}
At order $\omega^1 \kappa^1$ we are only interested in $v_y^{(1,1)}(x,\rho)$ \footnote{For the electric field in $x$-direction, the solution for $v_t^{(1,1)}$ and $v_x^{(1,1)}$ is still trivial, i.e. it vanishes. Note that the constraint is also satisfied.}. The relevant differential equation reads
\begin{equation}\label{eq:pde}
(1+3\rho^4) \, \partial_\rho v_y^{(1,1)} - \rho (1-\rho^4) \partial_\rho^2 v_y^{(1,1)} - \rho \partial_x^2 v_y^{(1,1)} + 8 i \, \rho^2 \, \partial_\rho A_z=0 \, .
\end{equation}
The solution to this differential equation is not $x$-independent and hence not simple to takle with analytic methods. Hence, we define the averaged (or integrated) current $\left\langle \bar{j}_y \right\rangle$ to be 
\begin{equation}
\left\langle \bar{j}_y \right\rangle = \frac{1}{L} \int_0^L dx  \, \left\langle j_y(x) \right\rangle \, .
\end{equation}
In order to determine $\left\langle \bar{j}_y \right\rangle$ we have to solve the partial differential equation \eqref{eq:pde}. Let us first solve the homogeneous equation, i.e. 
\begin{equation}
(1+3\rho^4) \, \partial_\rho v_y^{(1,1)} - \rho (1-\rho^4) \partial_\rho^2 v_y^{(1,1)} - \rho \partial_x^2 v_y^{(1,1)}=0 \, .
\end{equation}
It is convenient to use a separation of variables ansatz of the form
\begin{equation}\label{eq:sep}
v_y^{(1,1)}(x,\rho) = \sum\limits_{n} v_{y(n)}^{(1,1)}(\rho) f_n(x) \, . 
\end{equation}
The functions $f_n(x)$ are given by
\begin{equation}
f_n(x) = \exp\left(i \, \frac{2\pi n}{L} x \right)\, ,
\end{equation}
and hence the homogeneous differential equation for $v_{y(n)}^{(1,1)}(\rho)$ reads
\begin{equation}\label{eq:ode}
(1+3\rho^4) \, \partial_\rho v_{y(n)}^{(1,1)} - \rho (1-\rho^4) \partial_\rho^2 v_{y(n)}^{(1,1)} + \rho \left( \frac{2\pi n}{L} \right)^2 v_{y(n)}^{(1,1)}=0 \, .
\end{equation}
Note that this is now an ordinary differential equation. The averaged current $\left\langle \bar{j}_y \right\rangle$ can be read off from $v_y^{(1,1)}$, in particular (ignoring the Chern Simons term)
\begin{equation}
\left\langle j_y(x) \right\rangle =  \, \kappa\, \omega \, \lim\limits_{\rho \rightarrow 0} \partial_\rho^2 v_y^{(1,1)}(x,\rho) 
\end{equation}
and for the integrated current
\begin{equation}
\left\langle \bar{j}_y \right\rangle = \kappa\,\omega \lim\limits_{\rho \rightarrow 0} \partial_\rho^2 v_{y(0)}^{(1,1)}(\rho) \, .
\end{equation}
Note that $v_{y(n)}^{(1,1)}$ with $n\neq 0$ drops out since their integral over $x$ vanishes. Hence we can set $n=0$ in the differential equation. 

Integrating the homogeneous differential equation \eqref{eq:ode} for $n=0$ is now straight forward. First of all, in this case we can introduce $\tilde{v}_{y(0)}^{(1,1)} = \partial_\rho  v_{y(0)}^{(1,1)}$ and integrate the (now first order ordinary) differential equation
\begin{equation}\label{eq:sol1}
\tilde{v}_{y(0)}^{(1,1)} = \frac{\tilde{C} \, \rho}{1-\rho^4} \, .
\end{equation}
Imposing the boundary condition $v_{y(0)}^{(1,1)}(0)=0$ at the conformal boundary, we obtain the following solution for $v_{y(0)}^{(1,1)}(\rho)$ by integration \eqref{eq:sol1}
\begin{equation}
v_{y(0)}^{(1,1)}(\rho)=\frac{\tilde{C}}{4} \left( \log(1+\rho^2)  - \log(1-\rho^2)   \right)
\end{equation}
Note that the solution is not regular at the horizon.\\

Now, let us proceed with the inhomogeneous differential equation. Also using the separation ansatz \eqref{eq:sep} we end up with
\begin{equation}\label{eq:ode2}
(1+3\rho^4) \, \partial_\rho v_{y(n)}^{(1,1)} - \rho (1-\rho^4) \partial_\rho^2 v_{y(n)}^{(1,1)} + \rho \left( \frac{2\pi n}{L} \right)^2 v_{y(n)}^{(1,1)} + 8 i \, \rho^2 \, \partial_\rho A_{z(n)}=0 \, .
\end{equation}
Again, we only have to solve the differential equation for $n=0$ and we may view it as a first order differential equation for $\tilde{v}_{y(0)}^{(1,1)}$.
\begin{equation}\label{eq:ode3}
(1+3\rho^4) \, \tilde{v}_{y(0)}^{(1,1)} - \rho (1-\rho^4) \partial_\rho \tilde{v}_{y(0)}^{(1,1)}+ 8 i \, \rho^2 \, \partial_\rho A_{z(0)}=0 \, .
\end{equation}
We use the \textit{variation of constants} method and we promote the parameter $\tilde{C}$ to a function of $\rho,$ denoted by $\tilde{C}(\rho)$. We can derive a differential equation for $\tilde{C}(\rho)$ by substituting the ansatz for  $\tilde{v}_{y(0)}^{(1,1)}$
\begin{equation}
v_{y(0)}^{(1,1)}(\rho) = \tilde{C}(\rho) \, \frac{\rho}{1-\rho^4}
\end{equation}
into \eqref{eq:ode3}. In particular we obtain
\begin{equation}
\tilde{C}'(\rho) = 8\, i\, \partial_\rho A_{z(0)}(\rho)
\end{equation}
which give rise to the solution
\begin{equation}
\tilde{C}(\rho) = 8 \,i\, \partial_\rho A_{z(0)}(\rho) \,-\, 8\, i\, \partial_\rho A_{z(0)}(1) \, .
\end{equation}
In particular we fixed the constant such that the total solution
\begin{equation}
\tilde{v}_{y(0)}^{(1,1)}(\rho) = \left( 8 \,i\, A_{z(0)}(\rho) \,-\, 8\, i\,  A_{z(0)}(1)    \right) \, \frac{\rho}{1-\rho^4}
\end{equation}
is regular at $\rho=1.$ The average current is hence given by (again ignoring Chern-Simons terms)
\begin{equation}
\left\langle \bar{j}_y \right\rangle = \frac{\kappa}{i} \lim\limits_{\rho \rightarrow 0} \partial_\rho^2 v_{y(0)}^{(1,1)}(\rho) = \frac{\kappa}{i} \lim\limits_{\rho \rightarrow 0} \partial_\rho \tilde{v}_{y(0)}^{(1,1)}(\rho) =  8 \,\kappa \left( A_{z(0)}(0) - A_{z(0)}(1) \right)  \, .
\end{equation}
As explained in next section, adding the Chern-Simons contribution in order to get the expression for the consistent current we obtain the final result used throghout the main text: 
\begin{equation}\label{eq:final}
\left\langle \bar{j}_y \right\rangle_{C.S.} = -\, 8\, i\,  \omega \,\kappa\, A_{z(0)}(\rho=1)\,\, \longrightarrow \,\, \bar{\sigma}\equiv \frac{1}{i\omega} \langle \bar{j}_y \rangle_{C.S.} = \,-\, 8\, \kappa\, A_{z(0)}(\rho=1) \, .
\end{equation}

\subsection{The Chern-Simons term and the consistent current}

The total conductivity that we are aiming for is that given by the two point function of the \textit{consistent} current. The expression analyzed in the previous section is, however, the conductivity for a covariant current, which is what one gets by just solving the e.o.m. The conductivity for the covariant current can be computed via Kubo formulae as well, by using a correlator of a covariant and a consistent current, which is obtained by varying the 1-point function of the covariant current w.r.t. the desired source. 
\begin{equation}
\langle  J^\mu_V\rangle_{cons}=\lim\limits_{\rho\rightarrow 0}\sqrt{-g}\left\{ H^{\rho\mu}  +4 \,\kappa\, \epsilon^{\rho\mu\beta\eta\sigma}A_\beta H_{\eta\sigma} \right\}
\end{equation}
Implicitly we have already dropped out the counter terms. This is not strictly correct since the first term on the r.h.s. is actually infinite. Nevertheless, this is enough for our purposes, since the counterterm is multiplied by a logarithm and therefore does not produce finite contributions. This implies that the CS term, which is finite, will not be affected by the counter terms. The $\sim \kappa$ term is precisely the term that has to be subtracted in order to obtain the covariant current. Our goal is now to determine what is the contribution to the consistent-consistent two point function stemming from this term. We first expand to 1st order in perturbations

\begin{equation}
\langle  J^\mu_V\rangle_{cons}=\lim\limits_{\rho\rightarrow0}\sqrt{-g}\left\{ h^{\rho\mu}  +4 \,\kappa\, \epsilon^{\rho\mu\beta\eta\sigma}(a_\beta H_{\eta\sigma}+A_\beta h_{\eta\sigma}  )\right\}
\end{equation}
and compute the 2-point function

\begin{align}
\langle J^\nu_VJ^\mu_V  \rangle_{cons}=&\,\delta_{v_{\nu0}}\langle  J^\mu_V\rangle_{cons}\\
=&\lim\limits_{\rho\rightarrow0}\sqrt{-g}\left\{ \frac{h^{\rho\mu}}{\delta{v_{\nu0}}}  +4 \,\kappa\, \epsilon^{\rho\mu\beta\eta\sigma}(\frac{\delta a_\beta}{\delta{v_{\nu0}}} H_{\eta\sigma}+A_\beta\frac{ \delta h_{\eta\sigma} }{\delta{v_{\nu0}}} )\right\}
\end{align}
In our concrete case $H_{\rho \sigma}=0$ (Moreover here we see that even with a chemical potential, although this would not hold, the variation is w.r.t. to a vector field implies that the contribution would be subleading) and the only non trivial background gauge field is $A_z$ so
\begin{eqnarray}
\langle J^y_VJ^x_V  \rangle_{cons}=\lim\limits_{\rho\rightarrow0}\sqrt{-g}\left\{ \frac{\delta h^{\rho x}}{\delta{v_{y 0}}}  +8 \,\kappa\, \epsilon^{\rho r x z\eta\sigma}A_z\frac{ \delta\partial_{\rho}v_{\sigma} }{\delta{v_{y 0}} }\right\}
\end{eqnarray}
since momentum in every direction except for $x$ is taken to be zero then
\begin{align}
\langle J^y_VJ^x_V  \rangle_{cons}=&\lim\limits_{\rho\rightarrow0}\sqrt{-g}\left\{ \frac{\delta h^{\rho x}}{\delta{v_{y 0}}}  +{8 \,\kappa\, \epsilon^{\rho r x z t y}A_z\frac{ \delta \partial_{t}v_{y} }{\delta{v_{y 0}}} }\right\}\\
=&\lim\limits_{\rho\rightarrow0}\sqrt{-g}\left\{ \frac{\delta h^{\rho x}}{\delta{v_{y 0}}}  -i\,\omega\,8\, \kappa\, \epsilon^{\rho r x z t y}A_z\frac{ \delta v_{y} }{\delta{v_{y 0}}}\right\}
\end{align}
So, the finite contribution of the $\sim \kappa$ term is 
\begin{eqnarray}
\langle J^y_VJ^x_V  \rangle_{cons}-\langle J^y_VJ^x_V  \rangle_{cov}= -\,i\,\omega\, 8\, \kappa\, \epsilon(\rho x z t y)A_z(x,0)
\end{eqnarray}
By looking at equation \eqref{eq:pde} we see that $\epsilon(ytx\rho z)=1=\epsilon(\rho x z t y)$ so finally
\begin{eqnarray}\label{eq:CS}
\langle J^y_VJ^x_V  \rangle_{cons}-\langle J^y_VJ^x_V  \rangle_{cov}= \,-\,i\,\omega\, 8 \,\kappa\, A_z(x,0)
\end{eqnarray}
Since the Kubo formula is 
\begin{equation}
\sigma = \lim\limits_{\omega\rightarrow 0} \frac{1}{i\,\omega}\langle J_V J_V \rangle 
\end{equation}
we finally find that indeed the consistent current is given by $\langle j \rangle\,-\,8\, \kappa\, A_z(x,0)$, recovering \eqref{eq:final} after taking the average in x direction. Note that we made no use of periodicity neither we took the average in any direction so \eqref{eq:CS} applies in any case.

\section{Numerical techniques}\label{app:numerical}

We solved the system of partial differential equations (\ref{eq:eqs}, \ref{eq:eqs2}) numerically
by means of a pseudospectral collocation method based on the FFTW
library \cite{FFTW05} and written in C++. More specifically, we expanded
the unknown functions in a Fourier basis along the boundary direction
and in a Chebyshev polynomial basis set in the bulk direction and
solved for the collocation points on a Fourier-Lobatto grid. We used
the Jacobian-free Newton-Raphson algorithm for solving the resulting
nonlinear algebraic system where we tackled the related linear problem
with the stabilized biconjugate gradient method (BiCGSTAB) as linear
solver. We preconditioned the iterative solver by a finite difference
approximation of the Jacobian whose inverse is computed with the SuperLU
library \cite{Li:2005:OSA:1089014.1089017}. The resulting computational
problem is very well suited for parallel data runs which are queued
on a small Linux cluster by employing the GNU parallel package \cite{Tange2011a}.

\begin{figure}
\noindent \begin{centering}
\includegraphics[scale=0.23]{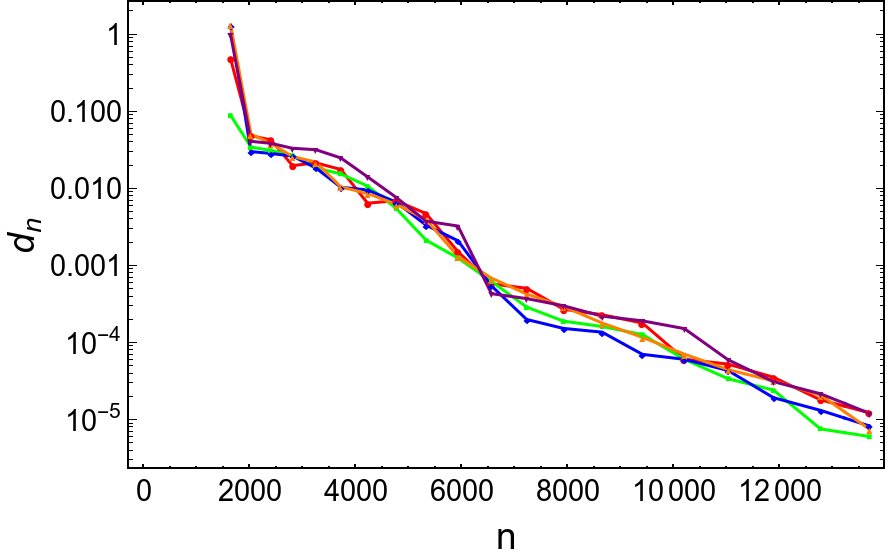}
\hspace{0.1cm}
\includegraphics[scale=0.23]{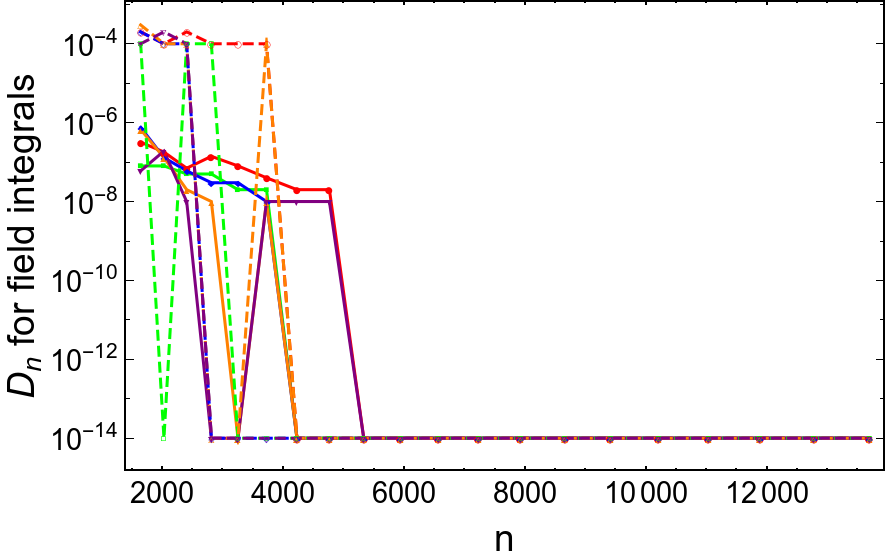}
\par\end{centering}
\caption{\label{fig:example-convergence-plot}\textbf{Left:} Example convergence plot for a generic random sample: maximum deviation from high resolution solution
$d_{n}$ as a function of the total number of grid points $n$. \textbf{Right:} Example convergence plot for a generic random sample: maximum deviation from high resolution solution
$d_{n}$ for integrals along the BH horizon (blue: gauge field integral,
orange: scalar field integral) as a function of the total number of
grid points $n$.}
\end{figure}

\subsection{Non-analyticity and logarithms}

The boundary expansion of the fields contains logarithmic terms which
effectively decrease the convergence rate of the Chebyshev expansion.
Given the boundary conditions in eqs. \eqref{eq:boundaryconditions} and taking the scalar field source to be a constant $M(x)=M$, as explained in the main text, we find the following expansions:
\begin{align}\label{eq:exp}
A_z(x,\rho) &= b(x) + \tilde{A_z}(x)\rho^2 + \left(b(x) M^2-b''(x)\right)\rho^2 \log(\rho)+\mathcal{O}(\rho^4, \rho^4\log (\rho))    \\
\phi(x,\rho) &= M \rho+ \tilde{\phi}(x)\rho^3 + \frac{1}{2}b(x)^2 M\rho^3 \log(\rho)+\mathcal{O}(\rho^5, \rho^5\log (\rho)) 
\end{align}
The lower the order where the first logarithmic term appears, the
worse the convergence rate. However, there are strategies to enhance
convergence in presence of logarithms. Here we use a simple ansatz
involving a coordinate transformation: 
\begin{align*}
\rho & =\zeta^{2},\qquad\textrm{with }\zeta\in\left[0,1\right]
\end{align*}
This effectively shifts logarithmic singularities of order $k$ at
the end points to order $2k$ and increases thus the convergence rate
(see \cite{BOYD198949}).

\subsection{Numerical quality and convergence studies}

Due to the overwhelmingly large number of solutions that need to be
found in order to improve the statistical estimates obtained from
the disorder model, it is not possible to save or inspect the convergence
for every sample, despite ensuring convergence in the Newton-Raphson
scheme. To check the numerical quality of the found solutions we studied
at least 3 samples with different generic random boundary data for
each tuple of parameters according to the following procedure: Given
a reference solution $u^{ref}$, computed on a high-resolution grid,
we use the following quantity as a measure for the convergence of
a sequence of solutions $u_{n}$ obtained with increasing resolution:
\begin{align}\label{eq:sup}
d_{n} & =\sup_{\left(\rho,x\right)\in\varSigma}\left|u^{ref}\left(\rho,x\right)-u_{n}\left(\rho,x\right)\right|
\end{align}

\begin{figure}
\noindent \begin{centering}
\includegraphics[scale=0.22]{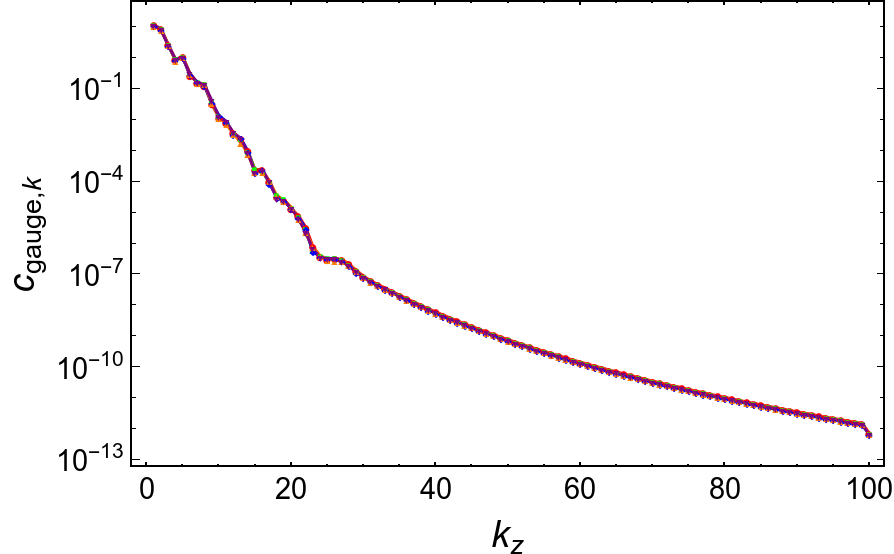}
\hspace{0.2cm}
\includegraphics[scale=0.22]{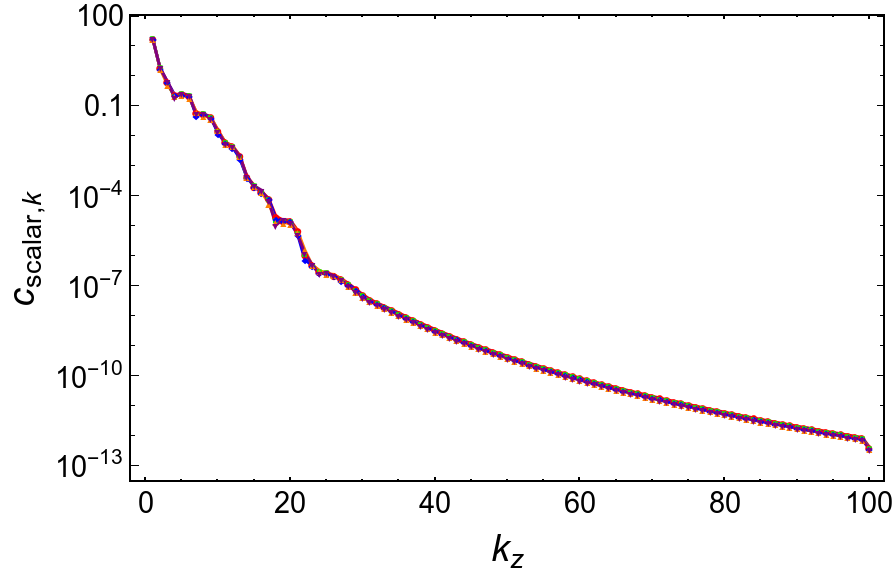}
\par\end{centering}
\caption{\label{fig:example-convergence-plot-2}
Example coefficient convergence
plot for the gauge field (left) and scalar field (right) for five randomly chosen samples at $M=10.1$,
$\bar\gamma=0.45$, $A_{M0}=0.1$, $b_{off}=1.0$ : maximum Chebyshev
absolute coefficient as a function wave number $k_{z}$, where the
maximum is taken over the remaining $x$-direction. }
\end{figure}

\begin{figure}
\noindent \begin{centering}
\includegraphics[scale=0.22]{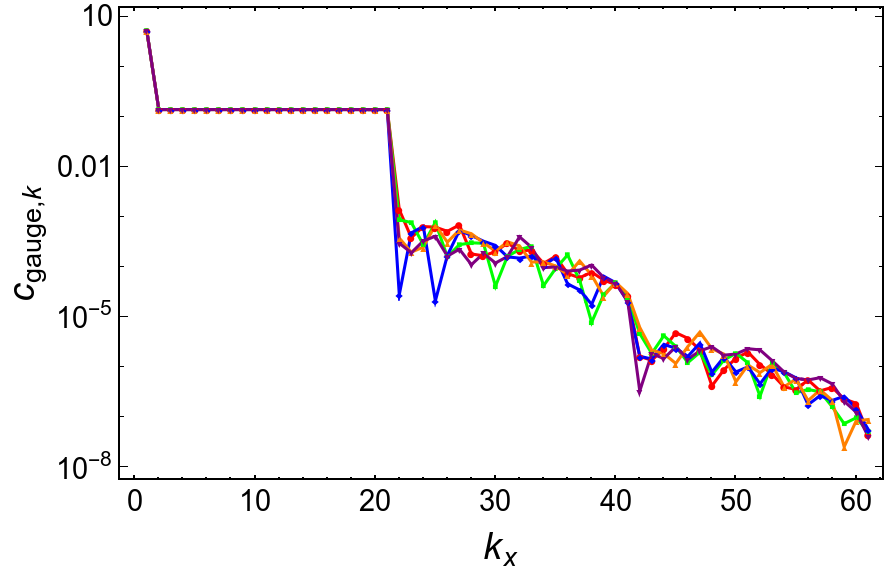}
\hspace{0.2cm}
\includegraphics[scale=0.22]{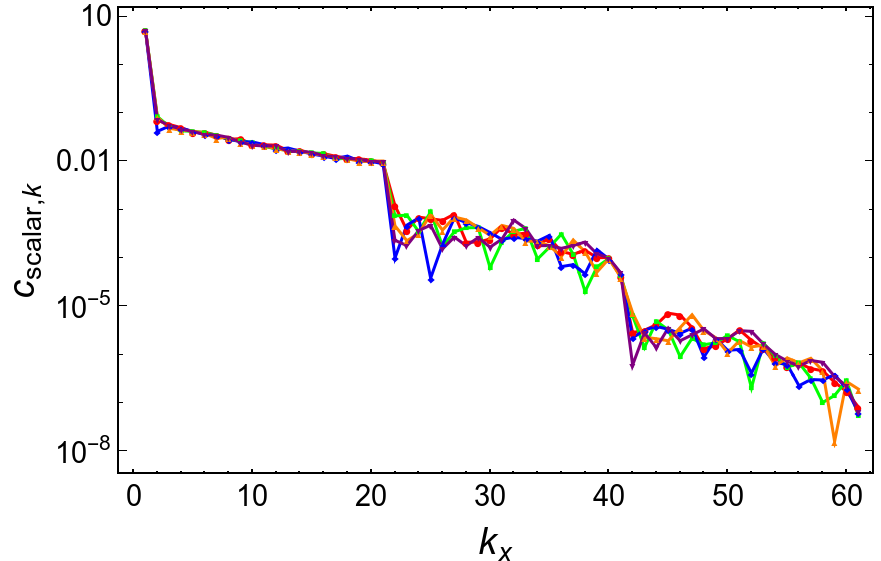}
\par\end{centering}
\caption{\label{fig:example-convergence-plot-2-2} Example coefficient convergence
plot for the gauge field (left) and scalar field (right) for five randomly chosen samples at $M=10.1$,
$\bar\gamma=0.45$, $A_{M0}=0.1$, $b_{off}=1.0$ : maximum absolute normalized
Fourier coefficient $c_{k}=\frac{1}{2}\sqrt{\alpha_{k}^{2}+\beta_{k}^{2}}$
(where $\alpha_{k}$ and $\beta_{k}$ are the coefficients of the
sine/cosine representation) as a function wave number $k_{x}$, where
the maximum is taken over the remaining $z$-direction.}
\end{figure}

We compute the supremum as follows: we interpolate the solutions $u^{ref}$
and $u_{n}$ on a high-resolution equidistant grid $\varSigma$ defined
on the $\left(\rho,x\right)$ domain. Using these interpolated solutions,
we compute the maximal deviation as in eq. \eqref{eq:sup}. Now monitoring the
convergence of $d_{n}$ with increasing resolution gives us the desired
convergence plot. Figure \ref{fig:example-convergence-plot} shows
the aforementioned convergence analysis for one generic random sample.
We see that the convergence is comparatively slow due to the high
number of random Fourier modes at the conformal boundary which couple
directly to the logarithmic contributions (cf. the boundary expansion
of the fields as specified in \eqref{eq:exp}). Nevertheless the convergence plot for the integrals
(see right panel of fig.~\ref{fig:example-convergence-plot}) shows a much faster
convergence, since we are here only extracting the very first Fourier
coefficient on the black hole horizon. So we can say that the numerical error is much smaller than the statistical error for all practical purposes.

In addition we studied the convergence of the spectral coefficients
for these samples in both directions, where we found convergence for
the used resolution (71x101) at least up to $10^{-7}$ in magnitude for
each direction (fig.~\ref{fig:example-convergence-plot-2}).
We clearly see the lag of convergence in the $z$-direction caused
by the logarithmic contributions (fig.~\ref{fig:example-convergence-plot-2}).
Furthermore we notice the plateau of Fourier coefficients for the
convergence plot regarding the $x$-direction, which originates from
the random coefficients of the boundary data (fig.~\ref{fig:example-convergence-plot-2-2}).

On top of that we studied the convergence of the extracted boundary
integrals in dependence of the resolution. The convergence speed of
the boundary integrals was even better than predicted by the supremum
based convergence estimate, which can be understood by noting that
these integrals are essentially the lowest Fourier coefficients of
the regarding spectral expansion. 
\bibliographystyle{JHEP-2}
\bibliography{DisWeyl}

\providecommand{\href}[2]{#2}\begingroup\raggedright\begin{thebibliography}{10}

\bibitem{sachdev2011}
S.~Sachdev, {\em Quantum phase transitions}.
\newblock Cambridge University Press, Cambridge, second ed.~ed., 2011.

\bibitem{2003RPPh...66.2069V}
M.~{Vojta}, {\it {Quantum phase transitions}},  {\em Reports on Progress in
  Physics} {\bf 66} (Dec., 2003) 2069--2110
  [\href{http://arXiv.org/abs/cond-mat/0309604}{{\tt cond-mat/0309604}}].

\bibitem{Bruin804}
J.~A.~N. Bruin, H.~Sakai, R.~S. Perry and A.~P. Mackenzie, {\it Similarity of
  scattering rates in metals showing t-linear resistivity},  {\em Science} {\bf
  339} (2013), no.~6121 804--807
  [\href{http://arXiv.org/abs/http://science.sciencemag.org/content/339/6121/804.full.pdf}{{\tt
  http://science.sciencemag.org/content/339/6121/804.full.pdf}}].

\bibitem{2007arXiv0709.0964V}
T.~{Vojta}, {\it {Computing quantum phase transitions}},  {\em ArXiv e-prints}
  (Sept., 2007) [\href{http://arXiv.org/abs/0709.0964}{{\tt 0709.0964}}].

\bibitem{2006JPhA...39R.143V}
T.~{Vojta}, {\it {TOPICAL REVIEW: Rare region effects at classical, quantum and
  nonequilibrium phase transitions}},  {\em Journal of Physics A Mathematical
  General} {\bf 39} (June, 2006) R143--R205
  [\href{http://arXiv.org/abs/cond-mat/0602312}{{\tt cond-mat/0602312}}].

\bibitem{2013AIPC.1550..188V}
T.~{Vojta}, {\it {Phases and phase transitions in disordered quantum systems}},
   in {\em American Institute of Physics Conference Series} (A.~{Avella} and
  F.~{Mancini}, eds.), vol.~1550 of {\em American Institute of Physics
  Conference Series}, pp.~188--247, Aug., 2013.
\newblock \href{http://arXiv.org/abs/1301.7746}{{\tt 1301.7746}}.

\bibitem{0022-3719-7-9-009}
A.~B. Harris, {\it Effect of random defects on the critical behaviour of ising
  models},  {\em Journal of Physics C: Solid State Physics} {\bf 7} (1974),
  no.~9 1671.

\bibitem{2013arXiv1309.0753V}
T.~{Vojta} and J.~A. {Hoyos}, {\it {Criticality and quenched disorder: rare
  regions vs. Harris criterion}},  {\em ArXiv e-prints} (Sept., 2013)
  [\href{http://arXiv.org/abs/1309.0753}{{\tt 1309.0753}}].

\bibitem{2004PSSBR.241.2118V}
T.~{Vojta} and R.~{Sknepnek}, {\it {Critical points and quenched disorder: From
  Harris criterion to rare regions and smearing}},  {\em Physica Status Solidi
  B Basic Research} {\bf 241} (July, 2004) 2118--2127
  [\href{http://arXiv.org/abs/cond-mat/0405070}{{\tt cond-mat/0405070}}].

\bibitem{2008PhRvL.100x0601H}
J.~A. {Hoyos} and T.~{Vojta}, {\it {Theory of Smeared Quantum Phase
  Transitions}},  {\em Physical Review Letters} {\bf 100} (June, 2008) 240601
  [\href{http://arXiv.org/abs/0802.2303}{{\tt 0802.2303}}].

\bibitem{PhysRevLett.90.107202}
T.~Vojta, {\it Disorder-induced rounding of certain quantum phase transitions},
   {\em Phys. Rev. Lett.} {\bf 90} (Mar, 2003) 107202.

\bibitem{PhysRevLett.23.17}
R.~B. Griffiths, {\it Nonanalytic behavior above the critical point in a random
  ising ferromagnet},  {\em Phys. Rev. Lett.} {\bf 23} (Jul, 1969) 17--19.

\bibitem{PhysRevLett.54.1321}
M.~Randeria, J.~P. Sethna and R.~G. Palmer, {\it Low-frequency relaxation in
  ising spin-glasses},  {\em Phys. Rev. Lett.} {\bf 54} (Mar, 1985) 1321--1324.

\bibitem{2010JLTP..161..299V}
T.~{Vojta}, {\it {Quantum Griffiths Effects and Smeared Phase Transitions in
  Metals: Theory and Experiment}},  {\em Journal of Low Temperature Physics}
  {\bf 161} (Oct., 2010) 299--323 [\href{http://arXiv.org/abs/1005.2707}{{\tt
  1005.2707}}].

\bibitem{0305-4470-36-43-017}
T.~Vojta, {\it Smearing of the phase transition in ising systems with planar
  defects},  {\em Journal of Physics A: Mathematical and General} {\bf 36}
  (2003), no.~43 10921.

\bibitem{PhysRevE.70.026108}
T.~Vojta, {\it Broadening of a nonequilibrium phase transition by extended
  structural defects},  {\em Phys. Rev. E} {\bf 70} (Aug, 2004) 026108.

\bibitem{PhysRevLett.108.185701}
L.~Demk\'o, S.~Bord\'acs, T.~Vojta, D.~Nozadze, F.~Hrahsheh, C.~Svoboda,
  B.~D\'ora, H.~Yamada, M.~Kawasaki, Y.~Tokura and I.~K\'ezsm\'arki, {\it
  Disorder promotes ferromagnetism: Rounding of the quantum phase transition in
  ${\mathrm{sr}}_{1\ensuremath{-}x}{\mathrm{ca}}_{x}{\mathrm{ruo}}_{3}$},  {\em
  Phys. Rev. Lett.} {\bf 108} (May, 2012) 185701.

\bibitem{2013AIPC.1550..263N}
D.~{Nozadze}, C.~{Svoboda}, F.~{Hrahsheh} and T.~{Vojta}, {\it {Modification of
  smeared phase transitions by spatial disorder correlations}},  in {\em
  American Institute of Physics Conference Series} (A.~{Avella} and
  F.~{Mancini}, eds.), vol.~1550 of {\em American Institute of Physics
  Conference Series}, pp.~263--267, Aug., 2013.
\newblock \href{http://arXiv.org/abs/1212.5962}{{\tt 1212.5962}}.

\bibitem{0295-5075-97-2-20007}
C.~Svoboda, D.~Nozadze, F.~Hrahsheh and T.~Vojta, {\it Disorder correlations at
  smeared phase transitions},  {\em EPL (Europhysics Letters)} {\bf 97} (2012),
  no.~2 20007.

\bibitem{Hartnoll:2016apf}
S.~A. Hartnoll, A.~Lucas and S.~Sachdev, {\it {Holographic quantum matter}},
  \href{http://arXiv.org/abs/1612.07324}{{\tt 1612.07324}}.

\bibitem{Ammon:2015wua}
M.~Ammon and J.~Erdmenger, {\em {Gauge/gravity duality}}.
\newblock Cambridge University Press, 2015.

\bibitem{zaanen2015holographic}
J.~Zaanen, Y.~Liu, Y.~Sun and K.~Schalm, {\em Holographic Duality in Condensed
  Matter Physics}.
\newblock Cambridge University Press, 2015.

\bibitem{DHoker:2009mmn}
E.~D'Hoker and P.~Kraus, {\it {Magnetic Brane Solutions in AdS}},  {\em JHEP}
  {\bf 10} (2009) 088 [\href{http://arXiv.org/abs/0908.3875}{{\tt 0908.3875}}].

\bibitem{DHoker:2010onp}
E.~D'Hoker and P.~Kraus, {\it {Magnetic Field Induced Quantum Criticality via
  new Asymptotically $AdS_5$ Solutions}},  {\em Class. Quant. Grav.} {\bf 27}
  (2010) 215022 [\href{http://arXiv.org/abs/1006.2573}{{\tt 1006.2573}}].

\bibitem{Iqbal:2010eh}
N.~Iqbal, H.~Liu, M.~Mezei and Q.~Si, {\it {Quantum phase transitions in
  holographic models of magnetism and superconductors}},  {\em Phys. Rev.} {\bf
  D82} (2010) 045002 [\href{http://arXiv.org/abs/1003.0010}{{\tt 1003.0010}}].

\bibitem{Landsteiner:2015pdh}
K.~Landsteiner, Y.~Liu and Y.-W. Sun, {\it {Quantum phase transition between a
  topological and a trivial semimetal from holography}},  {\em Phys. Rev.
  Lett.} {\bf 116} (2016), no.~8 081602
  [\href{http://arXiv.org/abs/1511.05505}{{\tt 1511.05505}}].

\bibitem{Gubankova:2014iha}
E.~Gubankova, M.~Cubrovic and J.~Zaanen, {\it {Exciton-driven quantum phase
  transitions in holography}},  {\em Phys. Rev.} {\bf D92} (2015), no.~8 086004
  [\href{http://arXiv.org/abs/1412.2373}{{\tt 1412.2373}}].

\bibitem{Iqbal:2011aj}
N.~Iqbal, H.~Liu and M.~Mezei, {\it {Quantum phase transitions in semilocal
  quantum liquids}},  {\em Phys. Rev.} {\bf D91} (2015), no.~2 025024
  [\href{http://arXiv.org/abs/1108.0425}{{\tt 1108.0425}}].

\bibitem{Donos:2012js}
A.~Donos and S.~A. Hartnoll, {\it {Interaction-driven localization in
  holography}},  {\em Nature Phys.} {\bf 9} (2013) 649--655
  [\href{http://arXiv.org/abs/1212.2998}{{\tt 1212.2998}}].

\bibitem{Baggioli:2016oqk}
M.~Baggioli and O.~Pujolas, {\it {On holographic disorder-driven
  metal-insulator transitions}},  {\em JHEP} {\bf 01} (2017) 040
  [\href{http://arXiv.org/abs/1601.07897}{{\tt 1601.07897}}].

\bibitem{Baggioli:2016oju}
M.~Baggioli and O.~Pujolas, {\it {On Effective Holographic Mott Insulators}},
  {\em JHEP} {\bf 12} (2016) 107 [\href{http://arXiv.org/abs/1604.08915}{{\tt
  1604.08915}}].

\bibitem{2011arXiv1112.6166D}
V.~{Dobrosavljevic}, {\it {Introduction to Metal-Insulator Transitions}},  {\em
  ArXiv e-prints} (Dec., 2011) [\href{http://arXiv.org/abs/1112.6166}{{\tt
  1112.6166}}].

\bibitem{Landsteiner:2015lsa}
K.~Landsteiner and Y.~Liu, {\it {The holographic Weyl semi-metal}},  {\em Phys.
  Lett.} {\bf B753} (2016) 453--457
  [\href{http://arXiv.org/abs/1505.04772}{{\tt 1505.04772}}].

\bibitem{Xu613}
S.-Y. Xu, I.~Belopolski, N.~Alidoust, M.~Neupane, G.~Bian, C.~Zhang, R.~Sankar,
  G.~Chang, Z.~Yuan, C.-C. Lee, S.-M. Huang, H.~Zheng, J.~Ma, D.~S. Sanchez,
  B.~Wang, A.~Bansil, F.~Chou, P.~P. Shibayev, H.~Lin, S.~Jia and M.~Z. Hasan,
  {\it Discovery of a weyl fermion semimetal and topological fermi arcs},  {\em
  Science} {\bf 349} (2015), no.~6248 613--617
  [\href{http://arXiv.org/abs/http://science.sciencemag.org/content/349/6248/613.full.pdf}{{\tt
  http://science.sciencemag.org/content/349/6248/613.full.pdf}}].

\bibitem{Liu864}
Z.~K. Liu, B.~Zhou, Y.~Zhang, Z.~J. Wang, H.~M. Weng, D.~Prabhakaran, S.-K. Mo,
  Z.~X. Shen, Z.~Fang, X.~Dai, Z.~Hussain and Y.~L. Chen, {\it Discovery of a
  three-dimensional topological dirac semimetal, na3bi},  {\em Science} {\bf
  343} (2014), no.~6173 864--867
  [\href{http://arXiv.org/abs/http://science.sciencemag.org/content/343/6173/864.full.pdf}{{\tt
  http://science.sciencemag.org/content/343/6173/864.full.pdf}}].

\bibitem{Hosur:2013kxa}
P.~Hosur and X.~Qi, {\it {Recent developments in transport phenomena in Weyl
  semimetals}},  {\em Comptes Rendus Physique} {\bf 14} (2013) 857--870
  [\href{http://arXiv.org/abs/1309.4464}{{\tt 1309.4464}}].

\bibitem{2015Sci...349..622L}
L.~{Lu}, Z.~{Wang}, D.~{Ye}, L.~{Ran}, L.~{Fu}, J.~D. {Joannopoulos} and
  M.~{Solja{\v c}i{\'c}}, {\it {Experimental observation of Weyl points}},
  {\em Science} {\bf 349} (Aug., 2015) 622--624
  [\href{http://arXiv.org/abs/1502.03438}{{\tt 1502.03438}}].

\bibitem{2015PhRvX...5c1013L}
B.~Q. {Lv}, H.~M. {Weng}, B.~B. {Fu}, X.~P. {Wang}, H.~{Miao}, J.~{Ma},
  P.~{Richard}, X.~C. {Huang}, L.~X. {Zhao}, G.~F. {Chen}, Z.~{Fang}, X.~{Dai},
  T.~{Qian} and H.~{Ding}, {\it {Experimental Discovery of Weyl Semimetal
  TaAs}},  {\em Physical Review X} {\bf 5} (July, 2015) 031013
  [\href{http://arXiv.org/abs/1502.04684}{{\tt 1502.04684}}].

\bibitem{Kharzeev:2015znc}
D.~E. Kharzeev, J.~Liao, S.~A. Voloshin and G.~Wang, {\it {Chiral magnetic and
  vortical effects in high-energy nuclear collisions—A status report}},  {\em
  Prog. Part. Nucl. Phys.} {\bf 88} (2016) 1--28
  [\href{http://arXiv.org/abs/1511.04050}{{\tt 1511.04050}}].

\bibitem{Landsteiner:2016led}
K.~Landsteiner, {\it {Notes on Anomaly Induced Transport}},  {\em Acta Phys.
  Polon.} {\bf B47} (2016) 2617 [\href{http://arXiv.org/abs/1610.04413}{{\tt
  1610.04413}}].

\bibitem{Gooth:2017mbd}
J.~Gooth {\em et.~al.}, {\it {Experimental signatures of the mixed
  axial-gravitational anomaly in the Weyl semimetal NbP}},  {\em Nature} {\bf
  547} (2017) 324--327 [\href{http://arXiv.org/abs/1703.10682}{{\tt
  1703.10682}}].

\bibitem{Li:2014bha}
Q.~Li, D.~E. Kharzeev, C.~Zhang, Y.~Huang, I.~Pletikosic, A.~V. Fedorov, R.~D.
  Zhong, J.~A. Schneeloch, G.~D. Gu and T.~Valla, {\it {Observation of the
  chiral magnetic effect in ZrTe5}},  {\em Nature Phys.} {\bf 12} (2016)
  550--554 [\href{http://arXiv.org/abs/1412.6543}{{\tt 1412.6543}}].

\bibitem{PhysRevX.5.031023}
X.~Huang, L.~Zhao, Y.~Long, P.~Wang, D.~Chen, Z.~Yang, H.~Liang, M.~Xue,
  H.~Weng, Z.~Fang, X.~Dai and G.~Chen, {\it Observation of the
  chiral-anomaly-induced negative magnetoresistance in 3d weyl semimetal taas},
   {\em Phys. Rev. X} {\bf 5} (Aug, 2015) 031023.

\bibitem{nature1}
H.~Li, H.~He, H.-Z. Lu, H.~Zhang, H.~Liu, R.~Ma, Z.~Fan, S.~Shun-Qing and
  J.~Wang, {\it Negative magnetoresistance in dirac semimetal cd3as2},  {\em
  Nature Communications} {\bf 7:10301} (Aug, 2016).

\bibitem{Grushin:2012mt}
A.~G. Grushin, {\it {Consequences of a condensed matter realization of Lorentz
  violating QED in Weyl semi-metals}},  {\em Phys. Rev.} {\bf D86} (2012)
  045001 [\href{http://arXiv.org/abs/1205.3722}{{\tt 1205.3722}}].

\bibitem{1983PhLB..130..389N}
H.~B. {Nielsen} and M.~{Ninomiya}, {\it {The Adler-Bell-Jackiw anomaly and Weyl
  fermions in a crystal}},  {\em Physics Letters B} {\bf 130} (Nov., 1983)
  389--396.

\bibitem{Jackiw:1999qq}
R.~Jackiw, {\it {When radiative corrections are finite but undetermined}},
  {\em Int. J. Mod. Phys.} {\bf B14} (2000) 2011--2022
  [\href{http://arXiv.org/abs/hep-th/9903044}{{\tt hep-th/9903044}}].

\bibitem{Lucas:2016omy}
A.~Lucas, R.~A. Davison and S.~Sachdev, {\it {Hydrodynamic theory of
  thermoelectric transport and negative magnetoresistance in Weyl semimetals}},
   {\em Proc. Nat. Acad. Sci.} {\bf 113} (2016) 9463
  [\href{http://arXiv.org/abs/1604.08598}{{\tt 1604.08598}}].

\bibitem{PhysRevLett.114.257201}
A.~Altland and D.~Bagrets, {\it Effective field theory of the disordered weyl
  semimetal},  {\em Phys. Rev. Lett.} {\bf 114} (Jun, 2015) 257201.

\bibitem{PhysRevB.93.075113}
A.~Altland and D.~Bagrets, {\it Theory of the strongly disordered weyl
  semimetal},  {\em Phys. Rev. B} {\bf 93} (Feb, 2016) 075113.

\bibitem{2015PhRvL.115x6603C}
C.-Z. {Chen}, J.~{Song}, H.~{Jiang}, Q.-f. {Sun}, Z.~{Wang} and X.~C. {Xie},
  {\it {Disorder and Metal-Insulator Transitions in Weyl Semimetals}},  {\em
  Physical Review Letters} {\bf 115} (Dec., 2015) 246603
  [\href{http://arXiv.org/abs/1507.00128}{{\tt 1507.00128}}].

\bibitem{2015PhRvL.114t6602Z}
Y.~X. {Zhao} and Z.~D. {Wang}, {\it {Disordered Weyl Semimetals and Their
  Topological Family}},  {\em Physical Review Letters} {\bf 114} (May, 2015)
  206602 [\href{http://arXiv.org/abs/1412.7678}{{\tt 1412.7678}}].

\bibitem{Roy:2016amv}
B.~Roy, R.-J. Slager and V.~Juricic, {\it {Global Phase Diagram of a Dirty Weyl
  liquid and Emergent Superuniversality}},
  \href{http://arXiv.org/abs/1610.08973}{{\tt 1610.08973}}.

\bibitem{2017PhRvB..95a4204L}
T.~{Louvet}, D.~{Carpentier} and A.~A. {Fedorenko}, {\it {New quantum
  transition in Weyl semimetals with correlated disorder}},  {\em Phys. Review
  B} {\bf 95} (Jan., 2017) 014204 [\href{http://arXiv.org/abs/1609.08368}{{\tt
  1609.08368}}].

\bibitem{PhysRevB.94.115137}
B.~Roy and S.~Das~Sarma, {\it Quantum phases of interacting electrons in
  three-dimensional dirty dirac semimetals},  {\em Phys. Rev. B} {\bf 94} (Sep,
  2016) 115137.

\bibitem{2014PhRvL.113b6602S}
B.~{Sbierski}, G.~{Pohl}, E.~J. {Bergholtz} and P.~W. {Brouwer}, {\it {Quantum
  Transport of Disordered Weyl Semimetals at the Nodal Point}},  {\em Physical
  Review Letters} {\bf 113} (July, 2014) 026602
  [\href{http://arXiv.org/abs/1402.6653}{{\tt 1402.6653}}].

\bibitem{Landsteiner:2017lwm}
K.~Landsteiner, E.~Lopez and G.~Milans~del Bosch, {\it {Quenching the CME via
  the gravitational anomaly and holography}},
  \href{http://arXiv.org/abs/1709.08384}{{\tt 1709.08384}}.

\bibitem{Grignani:2016wyz}
G.~Grignani, A.~Marini, F.~Pena-Benitez and S.~Speziali, {\it {AC conductivity
  for a holographic Weyl Semimetal}},  {\em JHEP} {\bf 03} (2017) 125
  [\href{http://arXiv.org/abs/1612.00486}{{\tt 1612.00486}}].

\bibitem{Landsteiner:2016stv}
K.~Landsteiner, Y.~Liu and Y.-W. Sun, {\it {Odd viscosity in the quantum
  critical region of a holographic Weyl semimetal}},  {\em Phys. Rev. Lett.}
  {\bf 117} (2016), no.~8 081604 [\href{http://arXiv.org/abs/1604.01346}{{\tt
  1604.01346}}].

\bibitem{Copetti:2016ewq}
C.~Copetti, J.~Fernández-Pendás and K.~Landsteiner, {\it {Axial Hall effect
  and universality of holographic Weyl semi-metals}},  {\em JHEP} {\bf 02}
  (2017) 138 [\href{http://arXiv.org/abs/1611.08125}{{\tt 1611.08125}}].

\bibitem{Rogatko:2017svr}
M.~Rogatko and K.~I. Wysokinski, {\it {Holographic calcualtion of the
  magneto-transport coefficients in Dirac semimetals}},
  \href{http://arXiv.org/abs/1712.01608}{{\tt 1712.01608}}.

\bibitem{Jacobs:2015fiv}
V.~P.~J. Jacobs, P.~Betzios, U.~Gursoy and H.~T.~C. Stoof, {\it
  {Electromagnetic response of interacting Weyl semimetals}},  {\em Phys. Rev.}
  {\bf B93} (2016), no.~19 195104 [\href{http://arXiv.org/abs/1512.04883}{{\tt
  1512.04883}}].

\bibitem{Ammon:2016mwa}
M.~Ammon, M.~Heinrich, A.~Jiménez-Alba and S.~Moeckel, {\it {Surface States in
  Holographic Weyl Semimetals}},  {\em Phys. Rev. Lett.} {\bf 118} (2017),
  no.~20 201601 [\href{http://arXiv.org/abs/1612.00836}{{\tt 1612.00836}}].

\bibitem{Liu:2018bye}
Y.~Liu and Y.-W. Sun, {\it {Topological nodal line semimetals in holography}},
  \href{http://arXiv.org/abs/1801.09357}{{\tt 1801.09357}}.

\bibitem{Aharony:2015aea}
O.~Aharony, Z.~Komargodski and S.~Yankielowicz, {\it {Disorder in Large-N
  Theories}},  {\em JHEP} {\bf 04} (2016) 013
  [\href{http://arXiv.org/abs/1509.02547}{{\tt 1509.02547}}].

\bibitem{Vegh:2013sk}
D.~Vegh, {\it {Holography without translational symmetry}},
  \href{http://arXiv.org/abs/1301.0537}{{\tt 1301.0537}}.

\bibitem{Andrade:2013gsa}
T.~Andrade and B.~Withers, {\it {A simple holographic model of momentum
  relaxation}},  {\em JHEP} {\bf 05} (2014) 101
  [\href{http://arXiv.org/abs/1311.5157}{{\tt 1311.5157}}].

\bibitem{Baggioli:2014roa}
M.~Baggioli and O.~Pujolas, {\it {Electron-Phonon Interactions, Metal-Insulator
  Transitions, and Holographic Massive Gravity}},  {\em Phys. Rev. Lett.} {\bf
  114} (2015), no.~25 251602 [\href{http://arXiv.org/abs/1411.1003}{{\tt
  1411.1003}}].

\bibitem{Alberte:2015isw}
L.~Alberte, M.~Baggioli, A.~Khmelnitsky and O.~Pujolas, {\it {Solid Holography
  and Massive Gravity}},  {\em JHEP} {\bf 02} (2016) 114
  [\href{http://arXiv.org/abs/1510.09089}{{\tt 1510.09089}}].

\bibitem{Baggioli:2015gsa}
M.~Baggioli and D.~K. Brattan, {\it {Drag phenomena from holographic massive
  gravity}},  {\em Class. Quant. Grav.} {\bf 34} (2017), no.~1 015008
  [\href{http://arXiv.org/abs/1504.07635}{{\tt 1504.07635}}].

\bibitem{Lucas:2014sba}
A.~Lucas and S.~Sachdev, {\it {Conductivity of weakly disordered strange
  metals: from conformal to hyperscaling-violating regimes}},  {\em Nucl.
  Phys.} {\bf B892} (2015) 239--268 [\href{http://arXiv.org/abs/1411.3331}{{\tt
  1411.3331}}].

\bibitem{Lucas:2014zea}
A.~Lucas, S.~Sachdev and K.~Schalm, {\it {Scale-invariant
  hyperscaling-violating holographic theories and the resistivity of strange
  metals with random-field disorder}},  {\em Phys. Rev.} {\bf D89} (2014),
  no.~6 066018 [\href{http://arXiv.org/abs/1401.7993}{{\tt 1401.7993}}].

\bibitem{Lucas:2015lna}
A.~Lucas, {\it {Hydrodynamic transport in strongly coupled disordered quantum
  field theories}},  {\em New J. Phys.} {\bf 17} (2015), no.~11 113007
  [\href{http://arXiv.org/abs/1506.02662}{{\tt 1506.02662}}].

\bibitem{Donos:2014yya}
A.~Donos and J.~P. Gauntlett, {\it {The thermoelectric properties of
  inhomogeneous holographic lattices}},  {\em JHEP} {\bf 01} (2015) 035
  [\href{http://arXiv.org/abs/1409.6875}{{\tt 1409.6875}}].

\bibitem{Garcia-Garcia:2015crx}
A.~M. García-García and B.~Loureiro, {\it {Marginal and Irrelevant Disorder
  in Einstein-Maxwell backgrounds}},  {\em Phys. Rev.} {\bf D93} (2016), no.~6
  065025 [\href{http://arXiv.org/abs/1512.00194}{{\tt 1512.00194}}].

\bibitem{Andrade:2017lsc}
T.~Andrade, A.~M. García-García and B.~Loureiro, {\it {Coherence effects in
  disordered geometries with a field-theory dual}},
  \href{http://arXiv.org/abs/1711.10953}{{\tt 1711.10953}}.

\bibitem{Hartnoll:2014cua}
S.~A. Hartnoll and J.~E. Santos, {\it {Disordered horizons: Holography of
  randomly disordered fixed points}},  {\em Phys. Rev. Lett.} {\bf 112} (2014)
  231601 [\href{http://arXiv.org/abs/1402.0872}{{\tt 1402.0872}}].

\bibitem{Hartnoll:2014gaa}
S.~A. Hartnoll and J.~E. Santos, {\it {Cold planar horizons are floppy}},  {\em
  Phys. Rev.} {\bf D89} (2014), no.~12 126002
  [\href{http://arXiv.org/abs/1403.4612}{{\tt 1403.4612}}].

\bibitem{SORNETTE1998239}
D.~Sornette, {\it Discrete-scale invariance and complex dimensions},  {\em
  Physics Reports} {\bf 297} (1998), no.~5 239 -- 270.

\bibitem{Hartnoll:2015faa}
S.~A. Hartnoll, D.~M. Ramirez and J.~E. Santos, {\it {Emergent scale invariance
  of disordered horizons}},  {\em JHEP} {\bf 09} (2015) 160
  [\href{http://arXiv.org/abs/1504.03324}{{\tt 1504.03324}}].

\bibitem{Hartnoll:2015rza}
S.~A. Hartnoll, D.~M. Ramirez and J.~E. Santos, {\it {Thermal conductivity at a
  disordered quantum critical point}},  {\em JHEP} {\bf 04} (2016) 022
  [\href{http://arXiv.org/abs/1508.04435}{{\tt 1508.04435}}].

\bibitem{Balasubramanian:2013ux}
K.~Balasubramanian, {\it {Gravity duals of cyclic RG flows, with strings
  attached}},  \href{http://arXiv.org/abs/1301.6653}{{\tt 1301.6653}}.

\bibitem{Flory:2017mal}
M.~Flory, {\it {Discrete scale invariance in holography revisited}},
  \href{http://arXiv.org/abs/1711.03113}{{\tt 1711.03113}}.

\bibitem{Ryu:2011vq}
S.~Ryu, T.~Takayanagi and T.~Ugajin, {\it {Holographic Conductivity in
  Disordered Systems}},  {\em JHEP} {\bf 04} (2011) 115
  [\href{http://arXiv.org/abs/1103.6068}{{\tt 1103.6068}}].

\bibitem{Fujita:2008rs}
M.~Fujita, Y.~Hikida, S.~Ryu and T.~Takayanagi, {\it {Disordered Systems and
  the Replica Method in AdS/CFT}},  {\em JHEP} {\bf 12} (2008) 065
  [\href{http://arXiv.org/abs/0810.5394}{{\tt 0810.5394}}].

\bibitem{Arean:2015sqa}
D.~Arean, L.~A. Pando~Zayas, I.~S. Landea and A.~Scardicchio, {\it {Holographic
  disorder driven superconductor-metal transition}},  {\em Phys. Rev.} {\bf
  D94} (2016), no.~10 106003 [\href{http://arXiv.org/abs/1507.02280}{{\tt
  1507.02280}}].

\bibitem{Arean:2014oaa}
D.~Arean, A.~Farahi, L.~A. Pando~Zayas, I.~S. Landea and A.~Scardicchio, {\it
  {Holographic p-wave Superconductor with Disorder}},  {\em JHEP} {\bf 07}
  (2015) 046 [\href{http://arXiv.org/abs/1407.7526}{{\tt 1407.7526}}].

\bibitem{Arean:2013mta}
D.~Arean, A.~Farahi, L.~A. Pando~Zayas, I.~S. Landea and A.~Scardicchio, {\it
  {Holographic superconductor with disorder}},  {\em Phys. Rev.} {\bf D89}
  (2014), no.~10 106003 [\href{http://arXiv.org/abs/1308.1920}{{\tt
  1308.1920}}].

\bibitem{Shinozuka1991}
M.~Shinozuka and G.~Deodatis, {\it Simulation of stochastic processes by
  spectral representation},  {\em Applied Mechanics Reviews} {\bf 44} (Apr,
  1991) 191--204.

\bibitem{2011PhRvB..83v4402H}
F.~{Hrahsheh}, D.~{Nozadze} and T.~{Vojta}, {\it {Composition-tuned smeared
  phase transitions}},  {\em Phys. Review B} {\bf 83} (June, 2011) 224402
  [\href{http://arXiv.org/abs/1103.5439}{{\tt 1103.5439}}].

\bibitem{DHoker:2012rlj}
E.~D'Hoker and P.~Kraus, {\it {Quantum Criticality via Magnetic Branes}},  {\em
  Lect. Notes Phys.} {\bf 871} (2013) 469--502
  [\href{http://arXiv.org/abs/1208.1925}{{\tt 1208.1925}}].

\bibitem{FFTW05}
M.~Frigo and S.~G. Johnson, {\it The design and implementation of {FFTW3}},
  {\em Proceedings of the IEEE} {\bf 93} (2005), no.~2 216--231. Special issue
  on ``Program Generation, Optimization, and Platform Adaptation''.

\bibitem{Li:2005:OSA:1089014.1089017}
X.~S. Li, {\it An overview of superlu: Algorithms, implementation, and user
  interface},  {\em ACM Trans. Math. Softw.} {\bf 31} (Sept., 2005) 302--325.

\bibitem{Tange2011a}
O.~Tange, {\it Gnu parallel - the command-line power tool},  {\em ;login: The
  USENIX Magazine} {\bf 36} (Feb., 2011) 42--47.

\bibitem{BOYD198949}
J.~P. Boyd, {\it The asymptotic chebyshev coefficients for functions with
  logarithmic endpoint singularities: mappings and singular basis functions},
  {\em Applied Mathematics and Computation} {\bf 29} (1989), no.~1 49 -- 67.

\end{thebibliography}\endgroup

\end{document}